\documentclass[preprint,epsf]{aastex}
\input epsf

\newcommand{\kms}{km~s$^{-1}$}
\newcommand{\teff}{$T_{\rm eff}$}
\newcommand{\mbol}{$M_{\rm bol}$}
\newcommand{\aori}{$\alpha$ Ori}

\received{Nov 12, 1999}
\revised{}
\accepted{}
\ccc{}
\cpright{}{}

\slugcomment{ApJ in press}
\shorttitle{Stellar [Fe/H] at the Galactic Center}
\shortauthors{Ram\'{\i}rez et al.}

\begin{document}

\title{Stellar Iron Abundances at the Galactic Center}

\vspace{10mm}
\author{Solange V. Ram\'{\i}rez\altaffilmark{1,2}, 
K. Sellgren\altaffilmark{1,2}}
\affil{Department of Astronomy, The Ohio State University}

\author{John S. Carr\altaffilmark{2}}
\affil{Naval Research Laboratory}

\author{Suchitra C. Balachandran\altaffilmark{2}}
\affil{Department of Astronomy, University of Maryland}

\author{Robert Blum\altaffilmark{1,2}}
\affil{Cerro Tololo Inter-American Observatory}

\author{Donald M. Terndrup \& Adam Steed}
\affil{Department of Astronomy, The Ohio State University}

\vspace{10mm}
\altaffiltext{1}{Visiting Astronomer, Cerro Tololo Inter-American Observatory.
CTIO is operated by AURA, Inc.\ under contract to the National Science
Foundation.}
\altaffiltext{2}{Visiting Astronomer, Infrared Telescope Facility. 
IRTF is operated by the University of Hawaii under contract to the National
Aeronautics and Space Administration.}

\begin{abstract}
We present measurements of [Fe/H] for six M supergiant stars and three 
giant stars within 0.5 pc of the Galactic Center (GC) and one M supergiant star
within 30 pc of the GC. 
The results are based on high-resolution ($\lambda / \Delta \lambda =$ 
40,000) $K-$band spectra, taken with CSHELL at the NASA Infrared 
Telescope Facility. 
We determine the iron abundance by detailed abundance analysis,
performed with the spectral synthesis program MOOG.
The mean [Fe/H] of the GC stars is determined to be near solar, 
[Fe/H] = +0.12 $\pm$ 0.22. 
Our analysis is a {\it differential} analysis, as we have observed and applied 
the same analysis technique to eleven cool, luminous stars in the solar 
neighborhood with similar temperatures and luminosities as the GC stars.
The mean [Fe/H] of the solar neighborhood comparison stars, 
[Fe/H] = +0.03 $\pm$ 0.16, is similar to that of the GC stars.
The width of the GC [Fe/H] distribution is found to be narrower than the 
width of the [Fe/H] distribution of Baade's Window in the bulge but 
consistent with the width of the [Fe/H] distribution
of giant and supergiant stars in the solar neighborhood. 
\end{abstract}

\keywords{Galaxy: abundances, Galaxy: center, stars: abundances, 
stars: late-type}

\section{Introduction}

\citet{mor96} and \citet{ser96} have summarized the relative importance of
star formation, gas inflow, and gas outflow in the Galactic Center (GC).
Gas from the inner disk flows into the nucleus, perhaps driven by a bar 
potential \citep{sta91,mor96}, and some of the stellar mass-loss from the 
bulge may fall into the GC \citep{bli93,jen94}.
\citet{ser96} have proposed that this inflow of disk gas results in
sustained star formation in a ``central molecular zone" (CMZ) within a
radius from the GC of $R \leq$ 200 pc.
The GC is observed to be currently forming stars in the CMZ
at a rate of 0.3--0.6 M$_{\odot}$ yr$^{-1}$ \citep{gus89}. 
Some of the GC gas incorporated into stars, enriched by stellar nucleosynthesis,
will be returned to the GC interstellar medium by stellar mass-loss and 
supernovae.
X-ray observations of hot gas in the GC suggest that some gas could be driven
temporarily or permanently
from the GC by a galactic fountain or wind \citep{bli93,mor96}, however,
making it unclear how much enriched material is incorporated into following
generations of star formation.
The presence of strong magnetic fields in the GC also likely plays a role
in whether enriched gas is driven from or is retained within the GC.

Continued star formation in the central few hundred parsecs of the Galaxy may
lead to higher metallicities within the CMZ. 
Chemical abundances in the disks of spiral galaxies are observed to reach 
their highest values at the center \citep[see ][]{pag81,shi90}.  
H II regions, planetary nebulae, and OB associations in the 
Milky Way also show a radial metallicity gradient \citep{sha83,fic91,mac94,
sim95,vil96,rud97,aff97,sma97,gum98}. Chemical evolution models strive to 
explain these gradients by considering the relative star formation and gas 
infall/outflow rates, and the metal abundance of the gas compared to the stars 
\citep[see ][]{aud76,ran91}.

The metallicity of GC stars is needed in order to constrain models and 
understand the stellar processes in the central parsecs of our Galaxy.
Measurements of stellar metallicities in the center of the Milky Way
Galaxy, which is obscured by approximately 30 mag of extinction at $V$, are
only now beginning to be feasible through infrared studies.
\citet{car00} found a solar Fe abundance in IRS 7, an M supergiant at
a distance of $R = $ 0.2 pc from the GC, from a detailed abundance analysis 
of CSHELL ($\lambda / \Delta \lambda $ = 40,000) $K-$band spectra.
Our work analyzes similar data for several M supergiant and giant stars 
located in the central cluster, within 0.5 pc of the GC, and also one M 
supergiant star located in the Quintuplet cluster, 30 pc away from the GC.

\section{Observations}

\subsection{Sample Selection}

The GC stars were selected with three requirements. First, they have to 
be brighter than $K$ = 9.5, a limit set by the sensitivity of the instrument.
An hour of integration time is required for high signal-to-noise (S/N) 
spectra for a $K$ = 9 
star when observing with CSHELL at the NASA Infrared Telescope Facility (IRTF). 
Second, the stars should be cool luminous stars, to ensure we are observing
photospheric absorption lines rather than the stellar wind emission lines 
characteristic of hot luminous stars in the GC. Third, these cool luminous
stars should have little or no water absorption.
This is because published line lists for water (wavelengths, excitation 
potentials, dissociation potentials, oscillator strengths, and damping 
constants) are not yet adequate for detailed abundance analysis of 
high-resolution stellar spectra.
The amount of water absorption is determined from $H$ and $K-$band 
low-resolution spectra ($\lambda / \Delta \lambda \sim$ 500) taken with the 
IRS at CTIO \citep{ram99}. 
The selected sample consists of nine stars located in the central cluster
($R <$ 0.5 pc), and one star located in the Quintuplet cluster ($R$ = 30 pc). 
The central cluster stars are IRS 7, IRS 11, IRS 19, IRS 22, BSD 72, BSD 114, 
BSD 124, BSD 129, and BSD 140 
\citep[names and positions are given by ][]{blu96a}.
The star located in the Quintuplet cluster is VR5-7 \citep{mon94}.
Eleven cool, luminous stars in the solar neighborhood were selected from the 
literature to be observed
and analyzed the same way as the GC stars. These eleven stars have 
known abundances from detailed analysis in the optical \citep{luc82a,
luc82b,lam84,smi85,smi86,luc89} and in the infrared \citep{car00}. These 
cool, luminous stars in the solar neighborhood
were carefully selected to be in the same range of effective
temperature and surface gravity as the GC stars. This is very important 
in order to allow a differential analysis comparison, which will cancel 
possible systematic errors such as NLTE effects (see Sec. 4.2). 
The selected cool, luminous stars in the solar neighborhood are listed
in Table 1.

\subsection{Data Acquisition and Reduction}

The $K-$band high-resolution ($\lambda / \Delta \lambda =$ 40,000)
spectra used in the abundance analysis were taken with CSHELL
at the IRTF.
The IRTF is a 3--m telescope located at Mauna Kea Observatory in Hawaii, 
and CSHELL is the facility cryogenic infrared echelle spectrograph 
\citep{tok90}.
The observations were carried out in 1993 May, 1996 June, 1996 August,
1997 July, 1998 May, 1998 July, and 1999 June.
The detector for the 1993 observations was a 256$\times$256 NICMOS-3 HgCdTe 
array, which provided a spectral coverage of $\approx$ 1000 \kms ~or 73 \AA. 
The detector for the
rest of the observations was a 256$\times$256 SBRC InSb array, which gives a 
smaller coverage of $\approx$ 750 \kms ~or 55 \AA.
The slit used to achieve a resolution of $\lambda / \Delta \lambda =$ 40,000
is 0.5\arcsec ~wide, 30\arcsec ~long, with a pixel scale of 0.25\arcsec ~per
pixel for 1993 data and 0.20\arcsec ~per pixel for later data.
Our highest quality data were taken in 1998 and in 1999 when new 
order-separating circular variable filters were installed, providing 
spectra free of fringes.

Eight iron lines were selected by inspecting the high resolution atlas of 
cool stars \citep{wal96} and by analyzing synthetic spectra, which includes
atomic and molecular features (see details in Section 3.0 and in \citet{car00}).
Thousands of lines of molecules like CN, CO and ${\rm H_{2}O}$ are present 
throughout the infrared spectrum of cool stars.
The selected iron lines are almost free of known molecular contamination and
unidentified lines. The eight iron lines are observed using three grating 
settings of CSHELL. Atomic parameters of the eight iron lines used in the abundance analysis are listed in Table 2. 

Spectra were acquired at two different positions along the slit and reduced
separately, to determine any systematic effects like 
fringes. Several individual spectrum pairs were observed per star, until
the desired signal to noise ratio was reached. 
Stars of spectral type A or B were observed as close to the program star's
airmass as possible to correct for telluric absorption features and to remove
fringes. 
Such stars have no significant spectral features in the observed 
wavelength regions.

Image Reduction and Analysis Facility (IRAF\altaffilmark{3}) and VISTA
were used for data reduction.\altaffiltext{3}
{The IRAF software is distributed by the National Optical
Astronomy Observatories under contract with the National Science Foundation}
The reduction process began by flat fielding the individual spectra with 
calibration lamp flats, which were taken for every grating setting. 
Sky subtraction was performed by subtracting two consecutive spectra taken at 
different positions along the slit. 
Bad pixels were replaced by an interpolated value computed from neighboring
pixels in both the dispersion and perpendicular directions.
Individual spectra were extracted with a 5 pixel wide aperture using the 
APSUM package in IRAF. 
Extracted spectra observed at the same position along the slit were averaged to 
produce two spectra per star, one spectrum for each position along the slit. 
Spectra of the program stars were then divided by the appropriate A or B type
atmospheric standards, observed at the same position along the slit and 
reduced in the same way, to remove telluric absorption features and fringes. 
Each spectrum was then multiplied by a blackbody of the same temperature as
the A or B star to put the spectra on a relative flux density 
(${\rm F_{\lambda}}$) scale. 
The temperature of the A or B star was determined from its spectral type.
More than five telluric absorption lines were used to obtain wavelength 
solutions. The
wavelength calibration was performed using the tasks of VISTA.
Spectra were shifted in wavelength to correct for radial velocity differences,
by comparing the observed and synthetic spectra. The spectra were normalized
by a linear fit of continuum bands. These continuum bands were chosen by 
comparing 
the observed spectrum of \aori ~from the high-resolution atlas \citep
{wal96} with the synthetic spectrum of \aori. The bands selected are 
regions in the spectrum where the synthesis and the normalized observations 
of \aori ~are coincident and equal to unity.
We estimated the error per pixel as the difference of observations taken 
at two positions along the slit. We computed the mean signal to noise ratio
(S/N) per pixel by the average of the signal to noise ratio per pixel for all
spectra observed for that star. The mean S/N  
per pixel for each star is listed in Tables 1 and 3.

\section{Abundance Analysis}

The abundance analysis was done using a current version of the LTE 
spectral synthesis program MOOG \citep{sne73}. 
The program requires a line list giving the wavelength, excitation
potential, gf--values, and damping constants for all atomic and molecular
lines that contribute to the spectrum. The method used to construct the line
list is described in Sec. 3.1.
In addition, an input model atmosphere for the effective temperature
and surface gravity appropriate for each star and a value for the 
microturbulent velocity is also required.
The solar-abundance model atmospheres from Plez \citep{ple92a,ple92b} were 
used in our analysis because they are the only ones that include gravities and 
temperatures as low as those of the Galactic Center stars.
The Plez models also include sphericity, which is appropriate for supergiant
stars \citep{ple92b}. MOOG, however, does not account for sphericity in its
calculations. \citet{car00}, who also used MOOG, compute the effective 
temperature and
microturbulent velocity of \aori, VV Cep (HR8383), and $\beta$ And using 
different sets of model atmospheres, including the Plez models and two 
plane-parallel 
atmospheres. When the stellar parameters derived from each model were used 
self-consistenly to derive [Fe/H], the result for [Fe/H] was the same. 
The determination of the stellar parameters for our sample is discussed in
Sec. 3.2.

\subsection{Atomic and Molecular Parameters}

The line list was created the same way as \citet{car00}, using two main
steps: 1. The solar spectrum was used to determine the gf--values
and the damping constants. 2. Minor adjustments were made by comparison to the 
Arcturus spectrum \citep{wal96}. 
An initial line list was compiled using Fe I line positions and 
energy levels
from \citet{nav94} and the \citet{kur93} line lists for CN and other atomic
lines. The damping constants for all atomic lines were initially set to twice 
that of the Uns\"{o}ld approximation for van der Waals broadening \citep{hol91}.
A synthetic spectrum 
was generated for the Sun, using the \citet{kur93} solar model. The gf--values
and damping constants were adjusted to match the observed solar spectrum
\citep{liv91}, when needed, and unidentified lines were noted. Then, a 
synthetic spectrum was generated for Arcturus, using the model atmosphere,
stellar parameters, and abundances from \citet{pet93}. 
The solar line list provided a 
good match to the Arcturus spectrum. 
The gf--values of low excitation lines, not
observable in the Sun, had to be adjusted. The final Fe I atomic 
parameters are listed in Table 2. 
A synthetic spectrum of \aori, generated using the final line list,
is compared in Figure 1 to the observed spectrum from \citet{wal96}
in the region of each of our eight Fe I lines.
A Plez model atmosphere with the stellar parameters listed in Table 3,
and the CNO abundances from \citet{lam84} were used.

The depth of line formation was examined using contribution functions, 
which provide an indication of spectral line formation coming 
from different layers of a stellar model atmosphere \citep[see details in][] 
{edm69,sne73}. The determination of the depth of line formation 
is needed to make sure each line is formed in the range of opacity ($\tau$)
covered by our model atmosphere. 
The eight iron lines used in this abundance analysis are formed in the range of
--2.0 $<$ log $\tau < $ 0.3, where Plez model atmospheres cover a range of
--5.6 $<$ log $\tau < $ 2.6 .
Note that the Sc I lines in supergiant stars, which neither we nor \citet[][who
included hyperfine splitting calculations for Sc I lines]{car00}
are able to model correctly, are shown by the contribution function to be 
partially formed at optical depths outside the range of the Plez models.
We therefore approximately model Sc I lines by scaling the Fe I line profiles to
the correct depth and width to allow us to separate the one Fe I line at 
22399 \AA ~which is blended with a 
Sc I line at 22400 \AA. In the case of IRS 7 this procedure was not successful.

\subsection{Stellar Parameters}

\subsubsection{Effective Temperature}

The effective temperature (\teff) is a key parameter for any 
abundance analysis.
For the cool, luminous stars in the solar neighborhood, 
the effective temperatures were taken from 
the literature. Some effective temperatures come from 
spectroscopic measurements \citep{luc82a,luc82b,luc89,car00}. 
Other effective temperatures come from the calibration of $(V-K)$ colors vs. 
\teff, where \teff ~comes from angular diameters measured by lunar occultation
\citep{smi85,smi86}. 
\aori ~has an \teff ~determination from its angular diameter 
measured by lunar occultation \citep[\teff ~= 3605 $\pm$ 43 K,][]{dyc98},
which is very good agreement with the spectroscopic value obtained by 
\citet{car00} (\teff ~= 3540 $\pm$ 260, see Table 1).

For IRS 7, the brightest infrared source in the GC, \citet{car00} determined
\teff ~spectroscopically by requiring the carbon abundance derived from CO 
lines to be independent of the lower excitation potential. 
We adopt the stellar parameters for IRS 7 from \citet{car00}, which were
derived using the standard Plez grid of model atmospheres.
Their \teff ~analysis relied largely on the second overtone CO lines
present in the $H-$band (1.5$\mu$m - 1.8$\mu$m), because most of the 
first overtone CO lines present in the $K-$band are saturated. 

For the remaining nine stars in the GC the $H-$band spectra are too faint to be
observed with CSHELL at the IRTF, because of high extinction. 
Therefore CO cannot be used to compute the effective temperature.
For those nine GC stars, broad molecular bands present in $H$ and $K-$band low 
resolution spectra ($\lambda / \Delta \lambda \sim$ 500) are used to 
estimate \teff ~\citep{ram99}. 
These features are CO (2.3 $\mu$m) and ${\rm H_{2}O}$ 
(1.9 $\mu$m). The CO strength increases with decreasing \teff ~and
decreasing surface gravity (log $g$). The ${\rm H_{2}O}$ strength also 
increases with decreasing \teff, but increases
with increasing log $g$. These two features together, therefore, provide 
two-dimensional spectral classification \citep{bal73,kle86,blu96b}.
The low resolution spectra of GC stars were taken with the IRS at CTIO, 
used in cross-dispersed mode to acquire the $H$ and $K-$band simultaneously 
\citep{ram99}.
Nearby late-type stars with \teff ~determinations based directly or indirectly
on the lunar occultation technique 
\citep{smi85,smi86,smi90,fer90,mcw90,dyc98,ric98} 
were observed to calibrate our \teff ~determination. 
The low resolution spectra of cool, luminous stars stars with known \teff 
~were taken with the IRS at CTIO, and with TIFKAM at MDM Observatory 
\citep{ram99}.
The distinction between giants and supergiants is determined 
by the presence or weakness of ${\rm H_{2}O}$, respectively. 
The CO index is computed in the same way as in \citet{blu96b}. 
Once the distinction between giants and 
supergiants is established, \teff ~is determined from the relations 
shown in Figure 2, which are the best unweighted linear fit to the data.  
VR 5-7, IRS 19, IRS 22, BSD 72, BSD 124, and BSD 129 were classified as 
supergiant stars, and IRS 11, BSD 114 and BSD 140 were classified as giant 
stars. The luminosity class for GC stars is also listed in Table 3.
Our values for \teff ~for GC stars agree with the results from \citet{blu96b},
for the stars that are common to both samples. Our value for \teff ~from 
the CO index of IRS 7, 3400 $\pm$ 300 K, is also in good agreement with
the spectroscopic value of \teff ~from \citet{car00}, 3470 $\pm$ 250 K.
IRS 7 has a low carbon abundance (Carr et al. 2000, [C/H] = -0.8). The good
agreement between the \teff ~from CO index and the spectroscopic \teff 
~suggests that our derived values of \teff ~are not strongly affected by
moderate variations in the abundance of carbon.
The typical uncertainty in \teff ~computed from the CO index is 300 K
for supergiants and 280 K for giants. 
The \teff ~for cool, luminous stars in the solar neighborhood is listed in 
Table 1 and for GC stars is listed in Table 3. 

\subsubsection{Surface Gravity}

For the cool, luminous stars in the solar neighborhood, 
the surface gravity, $g$, from the literature was used
\citep{luc82a,luc82b,smi85,smi86,luc89,car00}.
For the GC stars, the surface gravity is determined from the relation:
\begin{equation}
{\rm log}~g = {\rm log}(M/M_{\odot}) + 4~{\rm log}
(T_{\rm eff}/T_{\rm eff_{\odot}}) - {\rm log}(L/L_{\odot}) - 
{\rm log}~g_{\odot}. 
\end{equation}
\teff ~has been already determined. The luminosity, $L$, is computed from 
\mbol, which is determined from $K$, $A_{K}$, the GC 
distance, and the bolometric correction, B.C. A GC distance of 8 kpc is 
assumed \citep{rei93}, $K$ and $A_{K}$ are from
\citet{blu96a}, and the B.C. is taken from \citet{eli85}. The final 
\mbol ~for the GC stars is listed in Table 3. The uncertainty in \mbol
~from the GC stars comes mainly from uncertainties in the extinction curve, 
$E(H-K)/A_{K}$
, and it estimated to be $\pm$0.4 \citep{car00}.  Once \mbol ~and 
\teff ~are determined, the stars are placed 
in the HR diagram, and masses, $M$, are obtained by over-plotting evolutionary 
tracks of varying initial masses. Solar metallicity evolutionary tracks 
\citep{sch92} are used to determine masses for supergiant stars. 
For the coolest stars, that fall outside the \teff ~range of \citet{sch92} 
evolutionary tracks, asymptotic giant branch (AGB) tracks from \citet{mar96} 
were used to estimate their masses.
Note that the stellar evolution models assume mass loss for the more massive 
stars, and the current mass of the star at its present age is assumed.
Twice solar metallicity evolutionary tracks \citep{sch93} gave the same result
for the derived masses. 
The value of log $g$ for GC stars is listed in Table 3. The uncertainty is
estimated considering the uncertainties in mass, effective temperature and
luminosity. 
Figure 3 shows the HR diagram of GC stars and also the cool, luminous stars 
in the solar neighborhood.
It is seen that the GC stars and the cool, luminous stars in the solar 
neighborhood occupy the same
place in the HR diagram, hence are very similar types of stars.

\subsubsection{Microturbulent Velocity}

The values for the microturbulent velocity ($\xi$) for the cool, luminous stars 
found in the literature show considerable variation among authors.
\citet{car00} found for VV Cep (HR 8383)
a much lower value ($\xi$ = 3.7 $\pm$ 0.2 \kms) than the one from 
\citet[][$\xi$ = 5.0 $\pm$ 0.5 \kms]{luc82b}. 
\citet{car00} used CO 
lines of the same excitation potential, for which the carbon abundance should 
be independent of the equivalent widths of the lines when $\xi$ is correct. 
\citet{luc82b} used the same principle, but used Fe I lines instead of CO lines.

Seven of our Fe I lines have similar excitation potential, and those lines 
were used to derive the microturbulence for the cool, luminous stars in the 
solar neighborhood. This was done in an iterative way.
An Fe abundance was derived from the synthetic model of cool, luminous stars in the solar neighborhood,
which included lines of Fe, CN, and other lines.
Using the derived Fe abundance, a synthetic model was computed containing only 
Fe lines, and an equivalent width was measured for the Fe lines.
This way the equivalent width is free from contributions from CN or other 
species.
Then, the derived Fe abundances from the synthesis were plotted vs. the 
Fe equivalent widths. This process was repeated with different
microturbulent velocities until the Fe abundance was independent of the Fe
equivalent width. The final equivalent widths of the Fe lines used in this
process are listed in Table 4. 
The obtained value of $\xi$ for cool, luminous stars in the solar neighborhood
is listed in Table 1. 
The uncertainty in the obtained microturbulent velocity comes 
from the uncertainty in the slope in the graph of Fe abundance vs. equivalent
width, based on the scatter of the data points. 
Our value of $\xi$ for \aori ~($\xi$ = 2.8 $\pm$ 0.2) is slightly different
from the value of $\xi$ from \citet{car00} ($\xi$ = 3.23 $\pm$ 0.15).
There is good agreement between
our values of $\xi$ and the ones given by \citet{smi85,smi86}, where
the mean difference is 0.15 \kms. Our values of $\xi$ are systematically 
lower than the values given by \citet{luc82b} and \citet{luc89} (HR 8383,
HD 202380, HD 163428, BD+59 594, HD 232766), where
the mean difference is 2.4 \kms. Our value of $\xi$ for HR 8726 is in good
agreement with the value given by \citet{luc82a}.

We do not have sufficient S/N for the GC stars to apply this technique for 
finding $\xi$. Instead, to get the microturbulent velocity for the GC stars, 
a relation between log $g$ and $\xi$ is used.
\citet{mcw90} showed that log $g$ and $\xi$ are related in G and K giants.
The values of $\xi$ and log $g$ for cool, luminous stars in the solar 
neighborhood, including the values 
from \citet{car00} and \citet{smi85,smi86,smi90}, 
are plotted to get a relation to apply to the GC stars. Figure 4 shows
this relation for stars with \teff ~and log $g$ similar to those of the
GC stars. An unweighted linear fit to the data gives :
\begin{equation}
\xi = ( 2.78 - 0.82 \times {\rm log}~g )~{\rm km~s^{-1}}.
\end{equation}

The uncertainty in the microturbulent velocity is estimated from the 
uncertainty of the
fit ($\pm$0.4 \kms) and the uncertainty from the surface gravity.
The value of $\xi$ for IRS 7 obtained by this fit, $\xi$ = 3.3 $\pm$ 0.4 \kms,
is in good agreement with 
the value of $\xi$ from \citet{car00}, $\xi$ = 3.30 $\pm$ 0.34 \kms, obtained 
using CO lines.

\subsubsection{Macroturbulent Velocity}

The synthetic spectra have to be convolved with a macroturbulent broadening
function \citep{gra92}, and with instrumental broadening, to match the line
widths of the observed spectrum. The instrumental broadening is a gaussian of
full width half maximum given by
the spectral resolution of the instrument (7 \kms). 
CN lines in two bands (21798 \AA ~-- 
21804 \AA; 22403 \AA ~-- 22409 \AA) were used to determine the macroturbulent 
velocity ($\zeta$) for the cool, luminous stars in the solar neighborhood. Several synthetic spectra 
were generated by MOOG using different values of $\zeta$ and N abundances. 
Then, 
$\chi^{2}$ was computed for each synthetic spectrum, and the best set of
N abundance and $\zeta$ was chosen for the minimum $\chi^{2}$. Note that 
this procedure is not measuring $\zeta$, but determining the best line
broadening that fits the data. Also, the N abundance obtained by this technique
is not real, because the 
C and O abundances have not been measured independently. Figure 5
shows the $\chi^{2}$ contour maps, one for each band, for HR 8726.
Figure 5 also shows the observed spectrum of HR 8726 and the synthetic
spectrum for the values of $\zeta$ and N abundance which give a minimum 
$\chi^{2}$ for both bands. The macroturbulent velocity for
the cool, luminous stars in the solar neighborhood is given in Table 1. 
The uncertainty in $\zeta$
is given by the standard deviation of the values of $\zeta$ derived for each
band. This uncertainty is more conservative than the uncertainty derived 
from $\chi^{2}$ statistics.

For the GC stars, the same method was used for IRS 7, VR 5-7, IRS 19, 
IRS 22, and IRS 11. For the remaining stars, BSD 72, BSD 114, BSD 124, BSD 129,
and BSD 140, only one band with CN lines was observed and the band did not 
have enough S/N to carry out this method. 
The values of $\zeta$ in Table 1 were used to derive a relationship between
log $g$ and $\zeta$ for supergiant stars. \citet{gra79} established a 
relationship between $\xi$ and $\zeta$ for giant stars. 
\citet{mcw90} shows that $\xi$ depends on log $g$, so by combining these two 
dependences we derive a relation between $\zeta$ and log $g$.
The values of $\zeta$ for supergiant stars measured by the 
broadening of the CN lines are plotted vs. log $g$ to get a relation to apply 
to the fainter GC supergiant stars. 
Figure 6 shows this relation. An unweighted linear fit to the data gives :
\begin{equation}
\zeta = ( 14.5 - 4.40 \times {\rm log}~g )~{\rm km~s^{-1}}.
\end{equation}
The uncertainty in the microturbulent velocity is estimated from the 
uncertainty of the
fit ($\pm$2.4 \kms) and the uncertainty in the surface gravity.
Equation (3) was applied to BSD 72, BSD 124, and BSD 129.

For giant stars, the range in log $g$ is too narrow to derive a relationship.
A mean value of $\zeta$ = 9.0 $\pm$ 3.0 \kms~is used for the GC giant stars.
This mean value is derived from values of $\zeta$ for solar neighborhood
giant stars and IRS 11 measured by the broadening of the CN lines, as 
given in Tables 1 and 3.
This mean value was adopted for BSD 114 and BSD 140.

\subsection{Uncertainties}

Our basic observable parameters, \teff ~and \mbol, have uncertainties that 
come from the technique used to compute them (sections 3.2.1 \& 3.2.2). 
The surface gravity uncertainty includes both the uncertainty in \teff ~and 
the uncertainty in \mbol ~(section 3.2.2). The microturbulent
velocity uncertainty includes both the uncertainties in log $g$ and the 
uncertainty of the fit used to calculate it (section 3.2.3). 
The same is true for the macroturbulent velocity (section 3.2.4).

It is important to know what are the effects of each of these uncertainties in 
the abundance determination. 
The uncertainty in [Fe/H] for each stellar parameter
has been estimated by varying one of the stellar parameters and computing the
difference in iron abundance. 
A typical uncertainty in \teff ~of $\pm$300 K results in an uncertainty of 
$\mp$ 0.08 dex in [Fe/H]. 
A typical uncertainty of $\pm$0.35 in log $g$ implies an uncertainty of 
$\pm$0.12 dex in [Fe/H]. 
A typical uncertainty of $\pm$0.4 \kms ~in $\xi$ causes an uncertainty of 
$\mp$ 0.09 dex in [Fe/H].
A typical uncertainty of $\pm$1.8 \kms ~in $\zeta$ translates to an uncertainty
of $\pm$0.13 dex in [Fe/H]. 
The iron abundance is not very sensitive to uncertainties in \teff ~and $\xi$,
but it is more sensitive to uncertainties in log $g$ and $\zeta$. 
In addition to the uncertainty from the stellar parameters, 
the standard error must also be considered for each star. 
The standard error comes from the scatter in [Fe/H]
as derived from individual Fe I lines. 
All the uncertainties from stellar parameters and the
standard error for each star are listed in Table 7 and 8. 
The total uncertainty is the quadratic addition of the uncertainties derived 
from the stellar parameters and the standard error in the value of [Fe/H]
from individual Fe I lines.
For the solar neighborhood stars and IRS 7, the quadratic addition of the
uncertainties is a good estimation of the total uncertainty, because all 
the uncertainties come from independent measurements.
In the cases where $\xi$ and $\zeta$ are computed though their relationships
with log $g$, the uncertainties are not independent, and the quadratic addition
can be an overestimation of the total uncertainty. 
We estimate that the total uncertainty will decrease by 
no more than 0.05 dex, if we consider the correlation of the uncertainties 
in log $g$, $\xi$ and $\zeta$.

Systematic errors may be present as well. They may come from unidentified
lines, which could make a contribution that is not included in our synthetic 
spectrum because of the lack
of atomic and molecular parameters. They may also come from errors in the
gf--values determination, or failure of the model atmospheres to correctly model
the stars we observe, because model atmospheres use solar abundance ratios, 
or NLTE effects. 

\subsection{NLTE effects}

The iron abundance could be affected by departures from LTE. The main NLTE 
effect in late-type stars is caused by overionization of electron donor 
metals by ultraviolet radiation \citep{aum75}. 
Using the Saha equation, we found that in our typical star 93 \% of iron
is neutral. Thus overionization should be smaller in our stars than in
warmer giants, where only a few percent of iron is neutral.

It is known from empirical studies that abundances derived from low excitation
lines give systematically lower abundances ($\sim$0.3 dex) than the ones 
derived from high excitation and ionized lines in late-type stars; high 
excitation and ionized lines give very similar abundance results
\citep{rul80,ste85,tak91,tom83}. 
Recently, \citet{the99} studied NLTE effects in Fe abundances in
metal-poor late-type stars. They found that ionized lines are not significantly
affected by NLTE, and that NLTE corrections become less important as [Fe/H]
increases, being minimal for solar abundance stars.
Mostly high excitation lines are used in our analysis, so the NLTE effects that
might be present should be minimal.

\citet{tom83} found that NLTE 
effects cancel out when a differential analysis is carried out relative to a 
very similar star in terms of effective temperature and luminosity.
Nearby stars of similar effective temperature and luminosity as
the GC stars (see Figure 2) have been analyzed similarly in this paper, 
providing a differential abundance comparison that should 
remove any NLTE effects that might be present. 

\section{Results}

Each iron abundance value for a particular Fe I line and star was derived from 
a comparison between the synthetic spectrum
generated by MOOG and the observed spectrum. The [Fe/H] results for each line
for cool, luminous stars in the solar neighborhood and GC stars are listed in 
Table 5 and 6, respectively. The uncertainties per line are estimated from the
ability to distinguish models with different [Fe/H] considering the S/N of
the observed spectrum. 
For cool, luminous stars in the solar neighborhood, the uncertainty per line 
is estimated to be $\pm$0.05 dex, 
since the S/N is very homogeneous among all lines and stars. For
GC stars, the uncertainty is listed individually for each line in
Table 6.
Figures 1, 7 and 8 show the synthetic and observed spectra
for each line for \aori, cool, luminous stars in the solar neighborhood, and GC
stars, respectively.

The final [Fe/H] results for each star are listed in Table 9. These values
are the mean [Fe/H] weighted by the individual uncertainties in [Fe/H] for
each Fe I line, as listed in Tables 5 and 6.
The final uncertainty in Table 9 is the total uncertainty from Table 7 and 8.

The analyses of our paper and \citet{car00} have five stars in common: IRS 7, 
\aori, HR 6146 (30 Her), HR 6702, and HR 8383 (VV Cep).
There is agreement in the Fe abundances within 0.1 dex for four stars 
(IRS 7, \aori, HR 6146, and HR 8383).
Our Fe abundance for HR 6702 is 0.18 dex lower than that of \citet{car00}.
The abundance technique used in both studies is very similar, 
but \citet{car00} included a different set of Fe I lines and slightly 
different stellar parameters. 

IRS 7 is known to be a variable supergiant \citep{blu96a}. 
There is one Fe line common to the analysis of \citet{car00} and of this
paper, which was observed at different epochs by \citet{car00} and by us.
For this line, our result is consistent within
the uncertainties with the results of \citet{car00}. 
Another spectrum of IRS 7, taken with NIRSPEC at the Keck Telescope one
year after our observationes was kindly made available to us 
(Figer, private communication). We have one Fe line in common with the
NISPEC spectrum, and again the iron abundance
results from both data sets are also consistent within the uncertainties.
We conclude that the effect of variability on our IRS 7 results are negligible
considering our uncertainties.

\subsection{Galactic Center Mean [Fe/H]}

An unweighted mean value of [Fe/H] = +0.09 is obtained for the GC stars. 
A mean weighted by the total uncertainties in [Fe/H] (Table 9) gives a value 
of [Fe/H] = +0.12 for the GC stars. The estimated dispersion or $\sigma$ 
is 0.22 dex, which is not significantly different from the typical total 
uncertainty in [Fe/H] for each GC star of 0.28 dex.
Even if the total uncertainties are overestimated by $\sim$0.05 dex by the 
quadratic addition of the individual uncertainties,
the typical uncertainty per star does not get small enough to resolve
the [Fe/H] distribution (see Sec. 3.3).
This means that the dispersion of the GC [Fe/H]
distribution can be understood by the uncertainties present in the data.

Note that the supergiant VR 5-7, at $R$ = 30 pc, has [Fe/H] = +0.09 $\pm$ 0.22, 
which is similar to the supergiants at $R <$ 0.5 pc, which have a mean 
[Fe/H] = +0.21 $\pm$ 0.15.

An unweighted mean of [Fe/H] = --0.05 and a mean weighted by the total 
uncertainties of [Fe/H] = +0.03 was obtained for the eleven 
cool, luminous stars in the solar neighborhood. 
The estimated dispersion or $\sigma$ is 0.16 dex, which again is not
significantly different from the typical total uncertainty in [Fe/H] 
which is 0.20 dex for these stars.

The obtained mean [Fe/H] for the GC is close to solar, and furthermore, 
very similar to the cool, luminous stars in the solar neighborhood that 
were analyzed in the same way.

\section{Discussion}

Our mean [Fe/H] for the GC stars, +0.12 $\pm$ 0.22, is similar to that of 
supergiants in the Galactic disk.
Figure 9a compares the GC [Fe/H] distribution with the [Fe/H] distribution
of 40 F, G, K, and M supergiants within 2 kpc of the Sun \citep{luc89}.
The mean [Fe/H] of the solar neighborhood supergiant stars from \citep{luc89}
is +0.13 with a dispersion of 0.20 dex.
The width and mean of the two distributions agree closely.
In Figure 9b, we compare the GC [Fe/H] distribution with that of the 11
cool supergiants and luminous giants we have analyzed in this paper.
Again, the mean and width of the [Fe/H] distribution for the GC stars and
solar neighborhood stars are consistent.
Unlike the \citet{luc89} sample of supergiants, our 11 nearby stars 
are restricted to the same range in stellar parameter space as the GC
sample; in addition, we have used the same analysis method, Fe I lines,
and stellar models for all the stars.
It is important to emphasize that this differential analysis makes the
abundance comparison between our samples of GC and nearby stars robust
against possible systematic errors that might effect the absolute values
of [Fe/H].
Hence, we can conclude that the Fe abundances for the sample of GC stars
in this paper are nearly identical to those of similar \teff ~and luminosity
stars in the solar neighborhood, with no evidence for a super-solar 
metallicity.

Our mean [Fe/H] for the GC stars of +0.12 $\pm$ 0.22 appears to be in conflict
with the iron abundance of the Pistol star, at $R =$ 30 pc in the Quintuplet
cluster. \citet{naj99} derive [Fe/H] $\sim$ 0.5 for the Pistol star.
They qualify their result as preliminary, since details of the
atmosphere modeling, such as the effects of charge exchange reactions,
have to be studied and included in future work.
The Pistol star abundance is derived from emission lines in the winds from
this hot star, so the atmospheric modeling techniques are fundamentally
different from those more established techniques we have used to study
abundances from photospheric absorption lines in cool giant or supergiant stars.

\citet{ser96} have proposed that the central cluster was built up by 
continuous, perhaps episodic, star formation over the lifetime of the 
Galaxy. In this case, the central 100 pc would form a stellar population
distinct from the old population of the Bulge.
In Figure 9c, the GC [Fe/H] distribution is compared with the [Fe/H]
distribution for 262 stars in Baade's Window (BW) in the bulge
\citep[$l,b=1^{\circ}, -4^{\circ}$; ][]{sad96}.
\citet{sad96} have a typical uncertainty per star of $\pm$0.2 dex, so their
finding that [Fe/H] ranges from --1.0 to +0.5 in BW shows the
abundance spread of BW's distribution is well-resolved by
their technique.
The distribution of [Fe/H] in the GC, however, has a similar width as the
uncertainties in [Fe/H], 0.28 dex per star. 
Stars as metal-rich or as metal-poor as observed in the Bulge are not found
in our sample of GC stars. However, our GC stars are restricted to a 
young to intermediate age population, and hence an old stellar population,
if it exists in the central 100 pc, is not sampled by the work of this 
paper.
A larger abundance spread could be represented in such an older population,
but a spectroscopic abundance analysis will require the sensitivity to
reach K and M giants. A more complete comparison must also include
measurements of the abundance patterns in the central regions relative
to the Bulge. For example, the Bulge K giants analyzed by \citet{mcw94}
show a distinct enhancement in the $\alpha-$elements, compared to Fe.

An IMF weighted toward more massive stars in the GC has been proposed by
\citet{mor96}. 
Recently, \citet{fig99} have found an IMF slanted towards massive stars for
the Arches and Quintuplet clusters, both located within 30 pc of the GC.
A history of chemical evolution dominated by massive stars is expected to
result in enhancements of oxygen and $\alpha-$elements, such as Mg, Si, Ca, 
and Ti, relative to Fe \citep{whe89}.
The next step of our GC project is to measure abundances of $\alpha-$elements 
and study selective enrichment in the GC. 

Chemical evolution models try to explain abundance patterns by 
considering the relative star formation rate, the gas infall and outflow rates,
the star formation history and the abundance of the gas compared to the stars 
\citep[see ][]{aud76,ran91}. 
Most chemical evolution models that reproduce the radial gradients in galaxies
are valid only for distances greater than 2 kpc from the Galactic Center 
\citep{tin78,sam97,chi99,por99}. 
Now that the molecular gas in the GC (from which stars form) has been 
extensively studied \citep{sta91,bli93,mor96}, and that the abundances of 
stars in the GC has been obtained 
for the first time, we strongly urge theoreticians to put all these data 
together into a detailed chemical evolution model describing the central 
parsecs of our Galaxy.

\section{Conclusions.}

We present the first measurement of stellar [Fe/H] for ten stars in the 
Galactic Center (GC), at distances from the GC of $R <$ 30 pc. 
Nine GC stars are located in the central cluster ($R <$ 0.5 pc) and one 
star is located in the Quintuplet cluster ($R$ = 30 pc). 
The abundance analysis is based on high-resolution 
($\lambda / \Delta \lambda =$ 40,000) $K-$band spectra.
The mean [Fe/H] of the GC is determined to be near solar, [Fe/H] = +0.12 
$\pm$ 0.22, and also similar to the mean [Fe/H] for cool, luminous stars 
in the solar neighborhood, [Fe/H] = +0.03 $\pm$ 0.16, observed and analyzed 
in the same way. 
The width of the GC [Fe/H] distribution, which ranges from [Fe/H] = --0.3
to +0.5, is found to be significantly narrower than the width of the [Fe/H]
distribution of Baade's Window, which ranges from [Fe/H] = --1.0 to +0.8.
The GC [Fe/H] distribution is consistent with the [Fe/H] distribution of 
supergiant stars in the solar neighborhood.
This suggest that the most luminous stars in the GC are unlikely to be 
dominated by bulge-like stars, and that the evolutionary path of the GC, 
while unique, is closer to the disk's than to the bulge's.
The Quintuplet star at $R$ = 30 pc has a similar [Fe/H] to stars located in 
the central cluster at $R <$ 0.5 pc.
In the future, abundance measurements of CNO and $\alpha-$elements are planned
to provide a complete view of the abundance patterns of the stars in the 
central regions of the Milky Way.

\acknowledgments
Support for this work was generously provided by the National Science Foundation
through NSF grant AST-9619230 to K.S. and R.D.B. S.V.R. also gratefully
acknowledges support from a Gemini Fellowship (grant \# GF-1003-97 from the 
Association of Universities for Research in Astronomy, Inc., under NSF 
cooperative agreement AST-8947990 and from Fundaci\'on Andes under project 
C-12984), from an Ohio State Presidential Fellowship, and from an Ohio
State Alumni Research Award.
S.C.B. acknowledges support from NSF grants AST-9618335 and AST-9819870, and
J.S.C. from the Office of Naval Research.
S.V.R. would like to thank E. Luck, R. Kraft, B. Plez, A. Pradhan, R. Pogge, 
A. Gould, A. Sills, and C. Sneden for useful suggestions and enlightening 
comments. We thank D. Figer for his generosity in sharing his spectrum of
IRS 7 in advance of publication. Electronic versions of our spectra are
available upon request to S.V.R.
\vspace{5mm}
\noindent


\clearpage

\begin{deluxetable}{llcccccc}
\tablewidth{0pt}
\tablecolumns{7}
\tablecaption{Stellar Parameters for Nearby Late Type Stars.\label{tab01}}
\tablehead{\colhead{Star} & \colhead{Spectral} & \colhead{S/N\tablenotemark{a}}&
\colhead{\teff\tablenotemark{a}} & \colhead{\mbol\tablenotemark{a}} & 
\colhead{log $g$ \tablenotemark{a}} & \colhead{$\xi$ \tablenotemark{a}} & 
\colhead{$\zeta$ \tablenotemark{a}} \\
\colhead{} & \colhead{Type} & \colhead{} &
\colhead{(K)} & \colhead{} &
\colhead{} & \colhead{(\kms)} & \colhead{(\kms)} } 
\startdata
HR 6146  &M6 III\tablenotemark{1}    &45\tablenotemark{2} 
&3250$\pm$100\tablenotemark{1}&
-5.5\tablenotemark{1}  & 0.2$\pm$0.3\tablenotemark{1}&
2.0$\pm$0.5\tablenotemark{2}& 9.6$\pm$1.1\tablenotemark{2}\\
HR 6702  &M5 II-III\tablenotemark{1} &36\tablenotemark{2}
&3300$\pm$100\tablenotemark{1}&
-3.4\tablenotemark{1}  & 0.7$\pm$0.3\tablenotemark{1}&
2.0$\pm$0.5\tablenotemark{2}& 7.8$\pm$1.8\tablenotemark{2}\\
HR 7442  &M5 IIIas\tablenotemark{3}  &36\tablenotemark{2}
&3450$\pm$100\tablenotemark{3}&
-4.0\tablenotemark{3}  & 0.5$\pm$0.3\tablenotemark{3}&
2.4$\pm$0.2\tablenotemark{2}&10.5$\pm$2.7\tablenotemark{2}\\
HR 8062  &M4 IIIas\tablenotemark{3}  &34\tablenotemark{2}
&3450$\pm$100\tablenotemark{3}&
-3.5\tablenotemark{3}  & 0.7$\pm$0.3\tablenotemark{3}&
1.7$\pm$0.2\tablenotemark{2}& 8.9$\pm$0.5\tablenotemark{2}\\
\aori    &M1-2 Iab-a\tablenotemark{4}&305\tablenotemark{5}
&3540$\pm$260\tablenotemark{4}&
-7.4\tablenotemark{6}  & 0.0$\pm$0.3\tablenotemark{6}&
2.8$\pm$0.2\tablenotemark{2}&14.7$\pm$0.5\tablenotemark{2}\\
HR 8383  &M2 Iape+\tablenotemark{7}  &100\tablenotemark{2}
&3480$\pm$250\tablenotemark{4}&
-6.8\tablenotemark{7,8}& 0.0$\pm$0.3\tablenotemark{7}&
2.7$\pm$0.2\tablenotemark{2}&14.4$\pm$2.1\tablenotemark{2}\\
HD 202380&M3 Ib\tablenotemark{7}     &46\tablenotemark{2}
&3600$\pm$200\tablenotemark{7}&
-5.7\tablenotemark{7,8}& 0.6$\pm$0.5\tablenotemark{7}&
2.5$\pm$0.2\tablenotemark{2}&16.2$\pm$0.5\tablenotemark{2}\\
HD 163428&K5 II\tablenotemark{7}     &67\tablenotemark{2}
&3800$\pm$200\tablenotemark{7}&
-5.5\tablenotemark{7,8}& 0.6$\pm$0.5\tablenotemark{7}&
2.3$\pm$0.2\tablenotemark{2}&11.5$\pm$3.2\tablenotemark{2}\\
BD+59 594&M1 Ib\tablenotemark{9}     &49\tablenotemark{2}
&4000$\pm$200\tablenotemark{9}&
-6.6\tablenotemark{8}  &-0.9$\pm$0.3\tablenotemark{9}&
3.1$\pm$0.2\tablenotemark{2}&18.2$\pm$1.0\tablenotemark{2}\\
HD 232766&M1 Iab\tablenotemark{9}    &38\tablenotemark{2}
&4000$\pm$200\tablenotemark{9}&
-6.6\tablenotemark{8}  & 0.2$\pm$0.3\tablenotemark{9}&
2.2$\pm$0.2\tablenotemark{2}&11.5$\pm$3.7\tablenotemark{2}\\
HR 8726  &K5 Iab\tablenotemark{10}    &57\tablenotemark{2}
&4000$\pm$200\tablenotemark{10}&
-5.2\tablenotemark{10,8}& 0.5$\pm$0.3\tablenotemark{10}&
2.4$\pm$0.2\tablenotemark{2}&11.4$\pm$0.5\tablenotemark{2} 
\enddata
\tablenotetext{a}{S/N = mean signal to noise ratio per pixel ; 
\teff~= effective temperature; 
\mbol ~= bolometric magnitude; $g$ = surface gravity; 
$\xi$ = microturbulent velocity; $\zeta$ = macroturbulent velocity.}
\tablerefs{ (1) \citet{smi85}; (2) this paper; (3) \citet{smi86};
(4) \citet{car00}; (5) \citet{wal96}; (6) \citet{lam84}; 
(7) \citet{luc82b}; (8) \cite{lan91};
(9) \citet{luc89}; (10) \citet{luc82a}.}
\end{deluxetable}

\clearpage

\begin{deluxetable}{ccc}
\tablewidth{0pt}
\tablecaption{Fe Line Data.\label{tab02}}
\tablehead{\colhead{Wavelength (\AA)} & \colhead{$\chi$ (eV) \tablenotemark{a}} 
& \colhead{log gf \tablenotemark{a}}}
\startdata
21781.82  &  3.415  & --4.485 \\
22381.27  &  5.844  & --1.458 \\
22386.90  &  5.033  & --0.481 \\
22391.22  &  5.320  & --1.600 \\
22398.98  &  5.099  & --1.249 \\
22818.82  &  5.792  & --1.296 \\
22838.60  &  5.099  & --1.325 \\
22852.17  &  5.828  & --0.612
\enddata
\tablenotetext{a}{$\chi$ = Excitation Potential; gf = gf--value determined in 
this paper.}
\end{deluxetable}

\clearpage

\begin{deluxetable}{lcccccccc}
\tablewidth{0pt}
\tablecolumns{7}
\tablecaption{Stellar Parameters for Galactic Center Stars.\label{tab03}}
\tablehead{ \colhead{Star} & \colhead{Luminosity} & 
\colhead{S/N\tablenotemark{a}} &
\colhead{\teff\tablenotemark{a}} & \colhead{\mbol\tablenotemark{a,b}} & 
\colhead{$M$\tablenotemark{a}} & \colhead{log $g$ \tablenotemark{a}} & 
\colhead{$\xi$ \tablenotemark{a}} & \colhead{$\zeta$ \tablenotemark{a}} \\
\colhead{} & \colhead{Class} & \colhead{} & \colhead{(K)} & \colhead{} 
& \colhead{(${\rm M_{\odot}}$)} & \colhead{} & \colhead{(\kms)} & 
\colhead{(\kms)} }
\startdata
IRS 7\tablenotemark{c}& I &52&3470$\pm$250&--9.0&17$\pm$3 &--0.6$\pm$0.2&
3.3$\pm$0.4& 20.6$\pm$2.7 \\
VR 5-7                & I &30&3500$\pm$300&--7.8&14$\pm$2 &--0.2$\pm$0.3&
2.9$\pm$0.5& 12.6$\pm$1.6 \\
IRS 19                & I &75&3650$\pm$300&--7.2&14$\pm$2 &  0.1$\pm$0.3&
2.7$\pm$0.5& 13.4$\pm$2.0 \\
IRS 22                & I &25&3550$\pm$300&--6.4&10$\pm$2 &  0.3$\pm$0.3&
2.5$\pm$0.5& 12.8$\pm$1.6 \\
BSD 124               & I &10&3600$\pm$300&--5.5& 7$\pm$3 &  0.4$\pm$0.3&
2.4$\pm$0.5& 12.7$\pm$2.7 \\
BSD 129               & I &13&3650$\pm$300&--5.3& 7$\pm$3 &  0.5$\pm$0.3&
2.4$\pm$0.5& 12.3$\pm$2.7 \\
BSD 72                & I &13&3750$\pm$300&--4.5& 5$\pm$2 &  0.8$\pm$0.3&
2.1$\pm$0.5& 11.0$\pm$2.7 \\
BSD 114               &III&15&3100$\pm$280&--5.8& 3$\pm$1 &--0.2$\pm$0.5&
2.9$\pm$0.6& 9.0$\pm$3.0 \\
IRS 11                &III&16&3100$\pm$280&--5.3& 3$\pm$1 &  0.0$\pm$0.5&
2.8$\pm$0.6& 9.1$\pm$2.2 \\
BSD 140               &III&13&3100$\pm$280&--4.8&2.5$\pm$1&--0.1$\pm$0.5&
2.9$\pm$0.6& 9.0$\pm$3.0
\enddata
\tablenotetext{a}{S/N = mean signal to noise ratio per pixel; 
\teff~= effective temperature; 
\mbol ~= bolometric magnitude; $M$ = mass; $g$ = surface gravity; 
$\xi$ = microturbulent velocity; $\zeta$ = macroturbulent velocity.}
\tablenotetext{b}{Error in \mbol ~is $\pm$0.4 \citep{car00}, 
dominated by the uncertainty in the extinction curve 
($ E(H-K) / A_{K}$ ).} 
\tablenotetext{c}{Stellar parameters from \citet{car00}.}
\end{deluxetable}

\clearpage

\begin{deluxetable}{lccccccc}
\tablewidth{0pt}
\tablecaption{Fe I Equivalent Widths for Nearby Late-Type Stars.\label{tab04}}
\tablehead{\colhead{} & \multicolumn{7}{c}{Wavelength of Fe I lines (\AA)} \\
\colhead{} & \colhead{22381.3} & \colhead{22386.9} & \colhead{22391.2} & 
\colhead{22399.0} & \colhead{22818.8} & \colhead{22838.6} & \colhead{22852.2} \\
\cline{1-8}
\colhead{Star} & \multicolumn{7}{c}{Equivalent Widths of Each Fe I line (m\AA)}}
\startdata
HR 6146    &  90$\pm$ 6 & 355$\pm$ 8 & 162$\pm$ 7 &
210$\pm$ 7 & 130$\pm$ 6 & 277$\pm$ 7 & 211$\pm$ 7 \\
HR 6702    &  67$\pm$ 5 & 332$\pm$ 9 & 136$\pm$ 7 &
189$\pm$ 7 & 119$\pm$ 6 & 234$\pm$ 7 & 220$\pm$ 7 \\
HR 7442    &  60$\pm$ 5 & 339$\pm$ 9 & 143$\pm$ 8 &
174$\pm$ 8 & 141$\pm$ 8 & 294$\pm$ 8 & 246$\pm$ 8 \\
HR 8062    &  65$\pm$ 5 & 351$\pm$ 9 & 149$\pm$ 7 &
199$\pm$ 6 & 110$\pm$ 6 & 208$\pm$ 7 & $-$        \\
\aori      & 101$\pm$ 8 & 520$\pm$10 & 218$\pm$10 &
319$\pm$10 & 178$\pm$10 & 339$\pm$10 & 271$\pm$10 \\
HD 202380  & 102$\pm$ 8 & 476$\pm$11 & 180$\pm$ 9 &
311$\pm$ 9 & 137$\pm$ 8 & 299$\pm$ 9 & $-$        \\
HR 8383    &  86$\pm$ 7 & 466$\pm$10 & 183$\pm$ 9 &
285$\pm$ 9 & 112$\pm$ 8 & 261$\pm$ 9 & 233$\pm$ 9 \\
HD 163428  &  87$\pm$ 7 & 430$\pm$ 9 & 172$\pm$ 9 &
241$\pm$ 9 & 113$\pm$ 8 & 263$\pm$ 8 & $-$        \\
BD+59 594  & 107$\pm$ 9 & 522$\pm$11 & 171$\pm$12 &
329$\pm$12 &  96$\pm$ 9 & 289$\pm$13 & 226$\pm$12 \\
HD 232766  &  66$\pm$ 6 & 410$\pm$ 8 & 159$\pm$ 9 &
228$\pm$ 9 &  97$\pm$ 8 & 216$\pm$ 9 & 173$\pm$ 9 \\
HR 8726    &  71$\pm$ 6 & 397$\pm$ 8 & 158$\pm$ 8 &
234$\pm$ 8 & 115$\pm$ 8 & 239$\pm$ 8 & 186$\pm$ 8
\enddata
\end{deluxetable}

\clearpage

\begin{deluxetable}{lcccccccc}
\tablewidth{0pt}
\tablecaption{[Fe/H] for Nearby Late Type Stars. \label{tab05}}
\tablehead{\colhead{} & \multicolumn{8}{c}{Wavelength of Fe I lines (\AA)} \\
\colhead{} & \colhead{21781.8} & \colhead{22381.3} &
\colhead{22386.9} & \colhead{22391.2} & \colhead{22399.0} &
\colhead{22818.8} & \colhead{22838.6} & \colhead{22852.2} \\
\cline{1-9}
\colhead{Star} & \multicolumn{8}{c}{[Fe/H] \tablenotemark{a} \ Determined from
Each Fe I Line}}
\startdata
HR 6146  & --0.05&  +0.05& --0.20&   0.00& --0.30&  +0.15&  +0.25&  +0.10 \\
HR 6702  & --0.05& --0.05& --0.25& --0.05& --0.30&  +0.20&  +0.10&  +0.30 \\
HR 7442  &  +0.10&  +0.10& --0.25&  +0.10& --0.60&  +0.20&  +0.20&  +0.20 \\
HR 8062  &  +0.10& --0.10& --0.05&  +0.05& --0.20&  +0.10& --0.05&   $-$  \\
\aori    &  +0.15&   0.00&  +0.05&  +0.05& --0.10&  +0.20&  +0.10&   0.00 \\
HD 202380&  +0.10&  +0.10&  +0.10&   0.00&  +0.10&  +0.10&  +0.10&   $-$  \\
HR 8383  & --0.05& --0.10& --0.10& --0.10& --0.20& --0.15& --0.25& --0.10 \\
HD 163428&  +0.07&   0.00&   0.00& --0.02& --0.25& --0.05& --0.05&   $-$  \\
BD+59 594& --0.10&   0.00& --0.25& --0.25& --0.25& --0.30& --0.35& --0.35 \\
HD 232766& --0.10& --0.20& --0.15& --0.15& --0.40& --0.20& --0.40& --0.40 \\
HR 8726  &  +0.15& --0.10&   0.00& --0.05& --0.20&   0.00& --0.10& --0.20 \\
\enddata
\tablenotetext{a}{Error in [Fe/H] for each line is $\pm$ 0.05 dex.}
\end{deluxetable}

\clearpage

\begin{deluxetable}{lcccccccc}
\tablewidth{0pt}
\tablecaption{[Fe/H] for Galactic Center Stars.\label{tab06}}
\tablehead{\colhead{} & \multicolumn{8}{c}{Wavelength of Fe I lines (\AA)} \\
\colhead{} & \colhead{21781.8} & \colhead{22381.3} &
\colhead{22386.9} & \colhead{22391.2} & \colhead{22399.0} &
\colhead{22818.8} & \colhead{22838.6} & \colhead{22852.2} \\
\cline{1-9}
\colhead{Star} & \multicolumn{8}{c}{[Fe/H] Determined from Each Fe I Line}}
\startdata
IRS 7    & +0.35& +0.22&--0.15&--0.10& --   & +0.38&--0.30& --   \\
 &$\pm$0.05&$\pm$0.07&$\pm$0.10&$\pm$0.05&         &$\pm$0.05&$\pm$0.05&        \\
VR 5-7   & +0.12& +0.20&--0.30& +0.15&--0.40& +0.10& +0.15& --   \\
 &$\pm$0.03&$\pm$0.07&$\pm$0.10&$\pm$0.05&$\pm$0.10&$\pm$0.05&$\pm$0.07&        \\
IRS 19   & +0.30& +0.40&--0.30& +0.40&--0.05& +0.45&--0.05& --   \\
 &$\pm$0.05&$\pm$0.05&$\pm$0.10&$\pm$0.07&$\pm$0.10&$\pm$0.05&$\pm$0.10&        \\
IRS 22   & +0.30&--0.05&--0.40& +0.15&--0.40& +0.20& +0.35& +0.30\\
&$\pm$0.05&$\pm$0.05&$\pm$0.10&$\pm$0.10&$\pm$0.10&$\pm$0.10&$\pm$0.10&$\pm$0.10\\
BSD 124  &  --  & +0.22& +0.20& +0.20&--0.10& +0.20& +0.20& --   \\
 &         &$\pm$0.07&$\pm$0.10&$\pm$0.20&$\pm$0.10&$\pm$0.05&$\pm$0.10&        \\
BSD 129  &  --  & +0.50& +0.50& +0.70& +0.20& +0.60& +0.60& --   \\
 &         &$\pm$0.20&$\pm$0.10&$\pm$0.20&$\pm$0.10&$\pm$0.10&$\pm$0.10&        \\
BSD 72   &  --  & +0.10& +0.40& +0.35& +0.05& +0.35&--0.20& --   \\
 &         &$\pm$0.10&$\pm$0.10&$\pm$0.10&$\pm$0.10&$\pm$0.10&$\pm$0.10&        \\
BSD 114  &  --  &--0.20&--0.85&--0.10&--0.70&~~0.00&--0.30& --   \\
 &         &$\pm$0.10&$\pm$0.20&$\pm$0.10&$\pm$0.10&$\pm$0.10&$\pm$0.10&        \\
IRS 11   &--0.15& +0.25&--0.80&--0.20&--0.80&--0.15&--0.50& --   \\
 &$\pm$0.05&$\pm$0.05&$\pm$0.10&$\pm$0.10&$\pm$0.10&$\pm$0.10&$\pm$0.10&        \\
BSD 140  &  --  &~~0.00&--0.40& +0.10&--0.40&~~0.00&~~0.00& --   \\   
 &         &$\pm$0.20&$\pm$0.20&$\pm$0.20&$\pm$0.20&$\pm$0.10&$\pm$0.10&    
\enddata
\end{deluxetable}

\clearpage

\begin{deluxetable}{lcccccc}
\tablewidth{0pt}
\tablecolumns{7}
\tablecaption{Uncertainties in [Fe/H] for Solar Neighborhood Stars.
\label{tab07}}
\tablehead{\colhead{Star} & \colhead{$\pm$ \teff \tablenotemark{a}} &
\colhead{$\pm$log $g$ \tablenotemark{a}} &
\colhead{$\pm \xi$ \tablenotemark{a}} &
\colhead{$\pm \zeta$ \tablenotemark{a}} &
\colhead{Std.\tablenotemark{b}} & 
\colhead{Total\tablenotemark{c}} }
\startdata
HR 6146   &$\mp$0.03 &$\pm$0.10 &$\mp$0.11 &$\pm$0.08 &$\pm$0.07 &$\pm$0.19\\
HR 6702   &$\mp$0.03 &$\pm$0.10 &$\mp$0.11 &$\pm$0.13 &$\pm$0.08 &$\pm$0.22\\
HR 7442   &$\mp$0.03 &$\pm$0.10 &$\mp$0.04 &$\pm$0.19 &$\pm$0.11 &$\pm$0.25\\
HR 8062   &$\mp$0.03 &$\pm$0.10 &$\mp$0.04 &$\pm$0.04 &$\pm$0.04 &$\pm$0.13\\
\aori     &$\mp$0.07 &$\pm$0.10 &$\mp$0.04 &$\pm$0.04 &$\pm$0.03 &$\pm$0.14\\
HD 202380 &$\mp$0.05 &$\pm$0.17 &$\mp$0.04 &$\pm$0.04 &$\pm$0.02 &$\pm$0.19\\
HR 8383   &$\mp$0.07 &$\pm$0.10 &$\mp$0.04 &$\pm$0.15 &$\pm$0.03 &$\pm$0.20\\
HD 163428 &$\mp$0.05 &$\pm$0.17 &$\mp$0.04 &$\pm$0.22 &$\pm$0.04 &$\pm$0.29\\
BD+59 594 &$\mp$0.05 &$\pm$0.10 &$\mp$0.04 &$\pm$0.07 &$\pm$0.05 &$\pm$0.15\\
HD 232766 &$\mp$0.05 &$\pm$0.10 &$\mp$0.04 &$\pm$0.26 &$\pm$0.05 &$\pm$0.29\\
HR 8726   &$\mp$0.05 &$\pm$0.10 &$\mp$0.04 &$\pm$0.04 &$\pm$0.04 &$\pm$0.13\\
\enddata
\tablenotetext{a}{\teff~= effective temperature; $g$ = surface gravity;
$\xi$ = microturbulent velocity; $\zeta$ = macroturbulent velocity.}
\tablenotetext{b}{standard error, determined from scatter among [Fe/H] values 
measured for different Fe I lines.}
\tablenotetext{c}{total uncertainty, derived from quadratic sum of 
uncertainties due to the standard error and the uncertainties from the 
stellar parameters (\teff, log $g$, $\xi$, $\zeta$).}
\end{deluxetable}

\clearpage

\begin{deluxetable}{lcccccc}
\tablewidth{0pt}
\tablecolumns{7}
\tablecaption{Uncertainties in [Fe/H] for Galactic Center Stars.
\label{tab08}}
\tablehead{\colhead{Star} & \colhead{$\pm$ \teff \tablenotemark{a}} &
\colhead{$\pm$log $g$ \tablenotemark{a}} &
\colhead{$\pm \xi$ \tablenotemark{a}} &
\colhead{$\pm \zeta$ \tablenotemark{a}} &
\colhead{Std.\tablenotemark{b}} &
\colhead{Total\tablenotemark{c}} }
\startdata
IRS 7     &$\mp$0.07 &$\pm$0.07 &$\mp$0.09 &$\pm$0.19 &$\pm$0.13 &$\pm$0.27\\
VR 5-7    &$\mp$0.08 &$\pm$0.10 &$\mp$0.11 &$\pm$0.11 &$\pm$0.11 &$\pm$0.23\\
IRS 19    &$\mp$0.08 &$\pm$0.10 &$\mp$0.11 &$\pm$0.14 &$\pm$0.13 &$\pm$0.26\\
IRS 22    &$\mp$0.08 &$\pm$0.10 &$\mp$0.11 &$\pm$0.11 &$\pm$0.12 &$\pm$0.24\\
BSD 124   &$\mp$0.08 &$\pm$0.10 &$\mp$0.11 &$\pm$0.19 &$\pm$0.06 &$\pm$0.26\\
BSD 129   &$\mp$0.08 &$\pm$0.10 &$\mp$0.11 &$\pm$0.19 &$\pm$0.08 &$\pm$0.27\\
BSD 72    &$\mp$0.08 &$\pm$0.10 &$\mp$0.11 &$\pm$0.19 &$\pm$0.10 &$\pm$0.27\\
BSD 114   &$\mp$0.08 &$\pm$0.17 &$\mp$0.13 &$\pm$0.21 &$\pm$0.16 &$\pm$0.35\\
IRS 11    &$\mp$0.08 &$\pm$0.17 &$\mp$0.13 &$\pm$0.15 &$\pm$0.18 &$\pm$0.33\\
BSD 140   &$\mp$0.08 &$\pm$0.17 &$\mp$0.13 &$\pm$0.21 &$\pm$0.10 &$\pm$0.32
\enddata
\tablenotetext{a}{\teff~= effective temperature; $g$ = surface gravity;
$\xi$ = microturbulent velocity; $\zeta$ = macroturbulent velocity.}
\tablenotetext{b}{standard error, determined from scatter among [Fe/H] values
measured for different Fe I lines.}
\tablenotetext{c}{total uncertainty, derived from quadratic sum of
uncertainties due to the standard error and the uncertainties from the
stellar parameters (\teff, log $g$, $\xi$, $\zeta$).}
\end{deluxetable}

\clearpage

\begin{deluxetable}{lccclcc}
\tablewidth{0pt}
\tablecolumns{5}
\tablecaption{Mean [Fe/H] for each star.\label{tab09}}
\tablehead{\multicolumn{3}{c}{Solar Neighborhood Stars} & \colhead{} &
\multicolumn{3}{c}{Galactic Center Stars} \\
\cline{1-3} \cline{5-7} 
\colhead{Star} & \colhead{N$_{\rm lines}$} & \colhead{[Fe/H]\tablenotemark{a}} 
& \colhead{} &
\colhead{Star} & \colhead{N$_{\rm lines}$} & \colhead{[Fe/H]\tablenotemark{a}} }
\startdata
HR 6146   & 8 &--0.01$\pm$0.19 & & IRS 7   & 6 &  +0.09$\pm$0.27 \\
HR 6702   & 8 &--0.02$\pm$0.22 & & VR 5-7\tablenotemark{b}&7&  +0.09$\pm$0.23 \\
HR 7442   & 8 & +0.01$\pm$0.25 & & IRS 19  & 7 &  +0.29$\pm$0.26 \\
HR 8062   & 7 &--0.03$\pm$0.13 & & IRS 22  & 8 &  +0.09$\pm$0.24 \\
\aori     & 8 & +0.05$\pm$0.14 & & BSD 124 & 6 &  +0.17$\pm$0.26 \\
HD 202380 & 7 & +0.07$\pm$0.19 & & BSD 129 & 6 &  +0.49$\pm$0.27 \\
HR 8383   & 8 &--0.14$\pm$0.20 & & BSD 72  & 6 &  +0.17$\pm$0.27 \\
HD 163428 & 7 &--0.05$\pm$0.29 & & BSD 114 & 6 & --0.29$\pm$0.35 \\
BD+59 594 & 8 &--0.24$\pm$0.15 & & IRS 11  & 7 & --0.16$\pm$0.33 \\
HD 232766 & 8 &--0.26$\pm$0.29 & & BSD 140 & 6 & --0.06$\pm$0.32 \\
HR 8726   & 8 &--0.07$\pm$0.13 & &         &   &                \\
\enddata
\tablenotetext{a}{Error in [Fe/H] is the total uncertainty from Table 7 and 8.}
\tablenotetext{b}{M supergiant star in the Quintuplet cluster ($R$ = 30 pc).}
\end{deluxetable}

\clearpage



\clearpage

\figcaption[]{Observed spectrum ($filled~squares$) of \aori ~from \citet{wal96}
compared with synthetic spectra generated by MOOG, using CNO abundances from 
\citet{lam84}, model atmosphere from \citet{ple92a}, and stellar parameters 
listed in Table 1.
The error bars come from the difference of observations taken in two epochs. 
We overplot a synthetic spectrum ($thick~line$) derived using the Fe abundances
listed in Table 5 and synthetic spectra ($thin~lines$) computed with Fe 
abundances different by $\pm$0.20 dex , which corresponds to the typical 
uncertainty in [Fe/H] for a solar neighborhood cool, luminous star.
Fe lines and their wavelengths in \AA ~are marked ($bold~vertical~lines$) at 
the top of each panel. 
CN, Sc, and unidentified lines ($question~marks$) are also marked 
($vertical~lines$). 
Tickmarks along the x-axis are 1 \AA ~apart.
The [Fe/H] value from Table 5 for each line is given in each panel.
\label{fig1}}

\figcaption[]{CO index vs effective temperature, \teff, for supergiant and 
giant stars. 
Upper panel: the calibration for supergiant stars, with
\teff ~from \citet{dyc98} ($filled~squares$) and \citet{ric98} ($open~circles$),
both measured from lunar occultations.
Lower panel: the calibration for giant stars.
Values of \teff ~from \citet{mcw90} ($open~triangles$), 
\citet{smi85,smi86,smi90} ($filled~circles$), and \citet{fer90} ($open~stars$)
are derived from a relationship between $V-K$ and \teff ~measured by lunar 
occultation techniques.
\teff ~from \citet{dyc98} ($open~circles$), are derived from lunar 
occultation measurements.
In both panels the uncertainties come from the literature.
The best unweighted linear fit to the data ($solid~lines$) is used to 
compute the \teff ~of the galactic center stars.
\label{fig2}}

\figcaption[]{HR diagram of galactic center (GC) stars ($filled~circles$) and 
cool, luminous stars in the solar neighborhood
($open~squares$). Stellar parameters and their uncertainties are given in 
Tables 1 and 3.
Solar metallicity evolutionary tracks \citep{sch92} are overplotted to 
determine masses for GC stars. 
The GC stars and the cool, luminous stars in the solar neighborhood occupy 
the same place in the HR diagram, hence are very similar types of stars.
\label{fig3}}

\figcaption[]{Relation between microturbulent velocity ($\xi$) and surface
gravity (log $g$). Values of $\xi$ and log $g$ are from Table 1 of this paper
($open~squares$), \citet{car00} ($crosses$), and \citet{smi85,smi86,smi90} 
($filled~triangles$).
The best unweighted linear fit to the data ($solid~line$) is used to compute 
$\xi$ for the GC stars. 
\label{fig4}}

\figcaption[]{Left panels: $\chi^{2}$ contour maps for determining 
the nitrogen abundance, log $\epsilon$ (N), and macroturbulent 
velocity ($\zeta$) for HR 8726.  
The minimum $\chi^{2}$ gives the best set of log $\epsilon$ (N) and $\zeta$
for each set of CN lines (21798 \AA ~-- 21804 \AA ~at the top, and 
22403 \AA ~-- 22409 \AA ~at the bottom). 
Right panels: observed spectrum of HR 8726 ($filled~squares$) with
error bars from the difference of observations taken at two positions along the
slit and the synthetic spectrum of HR 8726 ($solid~line$)
using the derived log $\epsilon$ (N) and $\zeta$ for each CN band
(21798 \AA ~-- 21804 \AA ~at the top, and 22403 \AA ~-- 22409 \AA ~at 
the bottom).
The synthetic spectrum is calculated by MOOG, using stellar parameters in
Table 1 and a model atmosphere from \citet{ple92a}.
The value of log $\epsilon$ (N) derived by this technique is not a true 
nitrogen abundance, because a combination of C and N abundances are needed
to derive a true value of log $\epsilon$ (N) from CN lines.
\label{fig5}}

\figcaption[]{Relation between macroturbulent velocity ($\zeta$) and surface
gravity (log $g$) for supergiant stars ($filled~circles$). 
All supergiant stars from Table 1 and IRS 7, VR 5-7, IRS 19, and IRS 22 from 
Table 3 are plotted. 
The best unweighted linear fit to the data ($solid~line$) is used to compute 
$\zeta$ for the GC supergiant stars BSD 72, BSD 124, and BSD 129.
\label{fig6}}

\figcaption[]{Observed ($filled~squares$) and synthetic spectra ($solid~line$)
for cool, luminous stars in the solar neighborhood.
The synthetic spectrum computed by MOOG, using the stellar
parameters listed in Table 1 and model atmospheres from \cite{ple92a}.
The error bars come from the difference of observations taken at two positions 
along the slit.
Fe lines and their wavelengths in \AA ~are marked ($bold~vertical~lines$) at
the top panels.
CN, Sc, and unidentified lines ($question~marks$) are also marked
($vertical~lines$).
Tickmarks along the x-axis are 1 \AA ~apart.
The [Fe/H] value from Table 5 for each line is given in each panel.
\label{fig7}}

\figcaption[]{Similar to Fig. 7, except these are observed and synthetic 
spectra for galactic center stars, using stellar parameters given in Table 3. 
Symbols are the same as Fig. 7.
The [Fe/H] value from Table 6 for each line is given in each panel.
\label{fig8}}

\figcaption[]{
(a) Fractional distribution of [Fe/H] for 10 GC stars (
$solid~line$) compared to [Fe/H] for 40 solar neighborhood supergiant stars
($dashed~line$) from \citet{luc89}. 
(b) Fractional distribution of [Fe/H] for
10 GC stars ($solid~line$) compared to [Fe/H] for 11 solar neighborhood stars
observed and analyzed in this paper ($dashed~line$).
(c) Fractional distribution of [Fe/H] for 10 GC stars ($solid$ 
$line$) compared to [Fe/H] for 262 Baade's window stars ($dashed$ $line$) 
from \citet{sad96}.
\label{fig9}}

\begin{figure}
\epsfxsize \hsize
\epsfbox{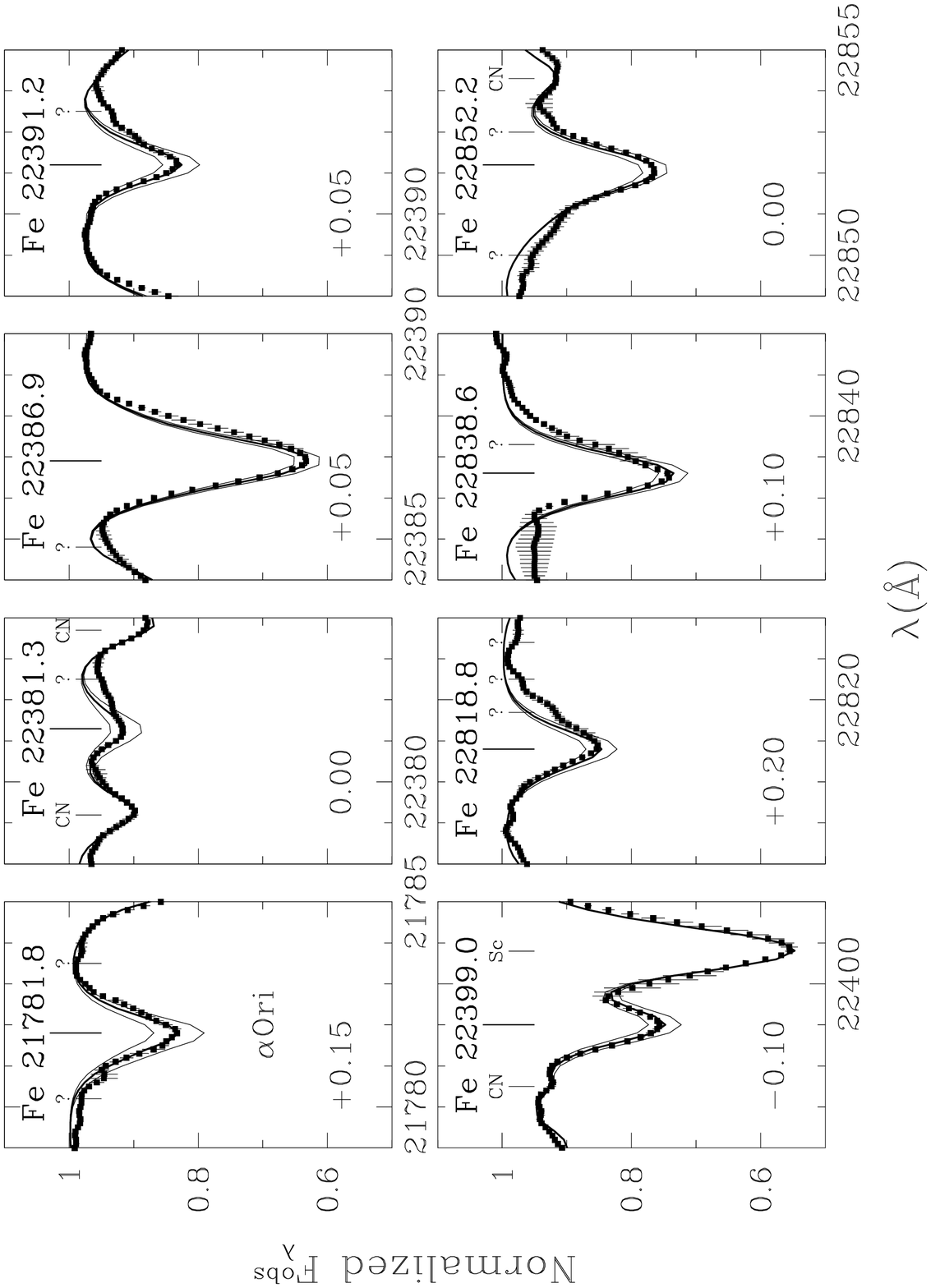}
\end{figure}

\begin{figure}
\epsfxsize \hsize
\epsfbox{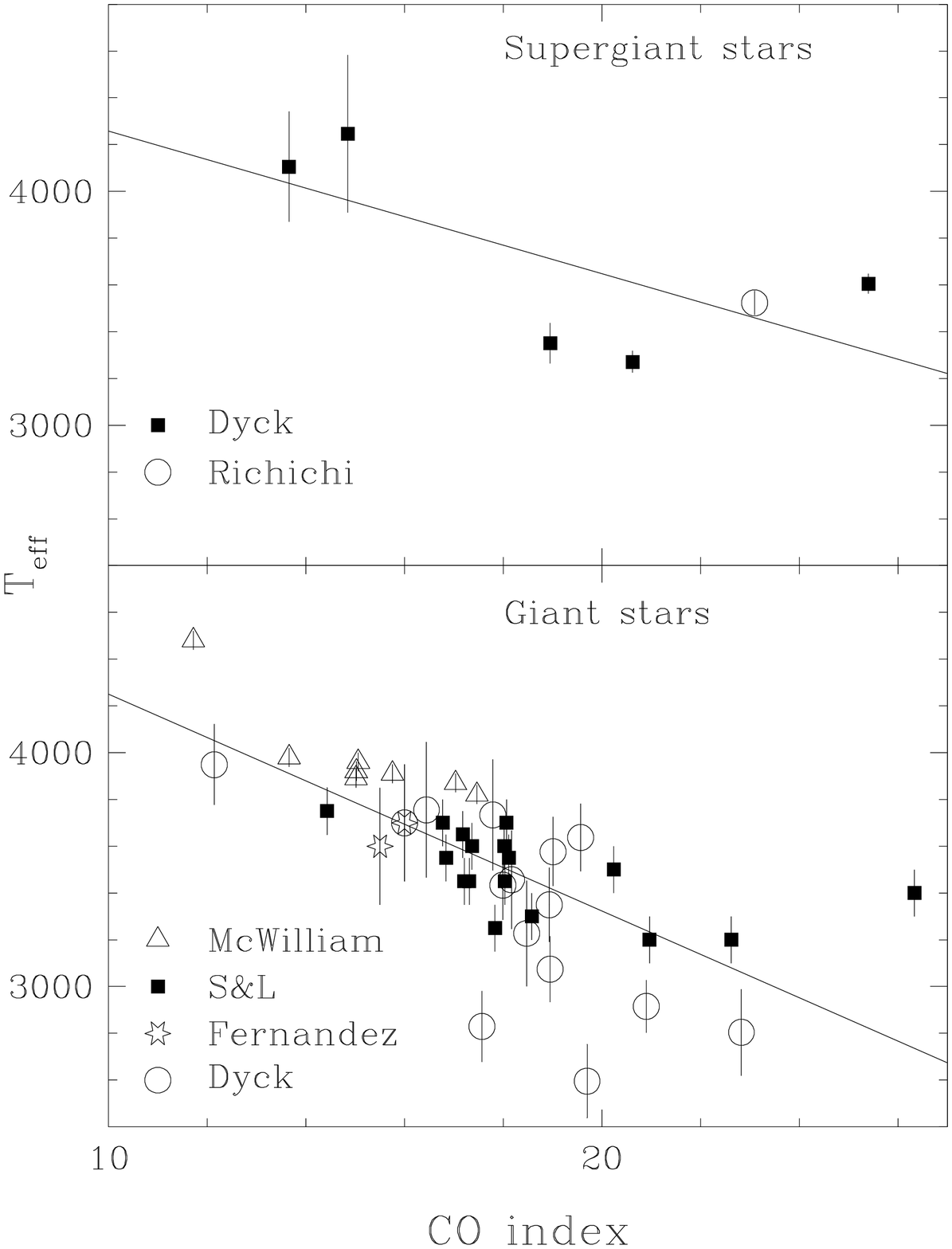}
\end{figure}

\begin{figure}
\epsfxsize \hsize
\epsfbox{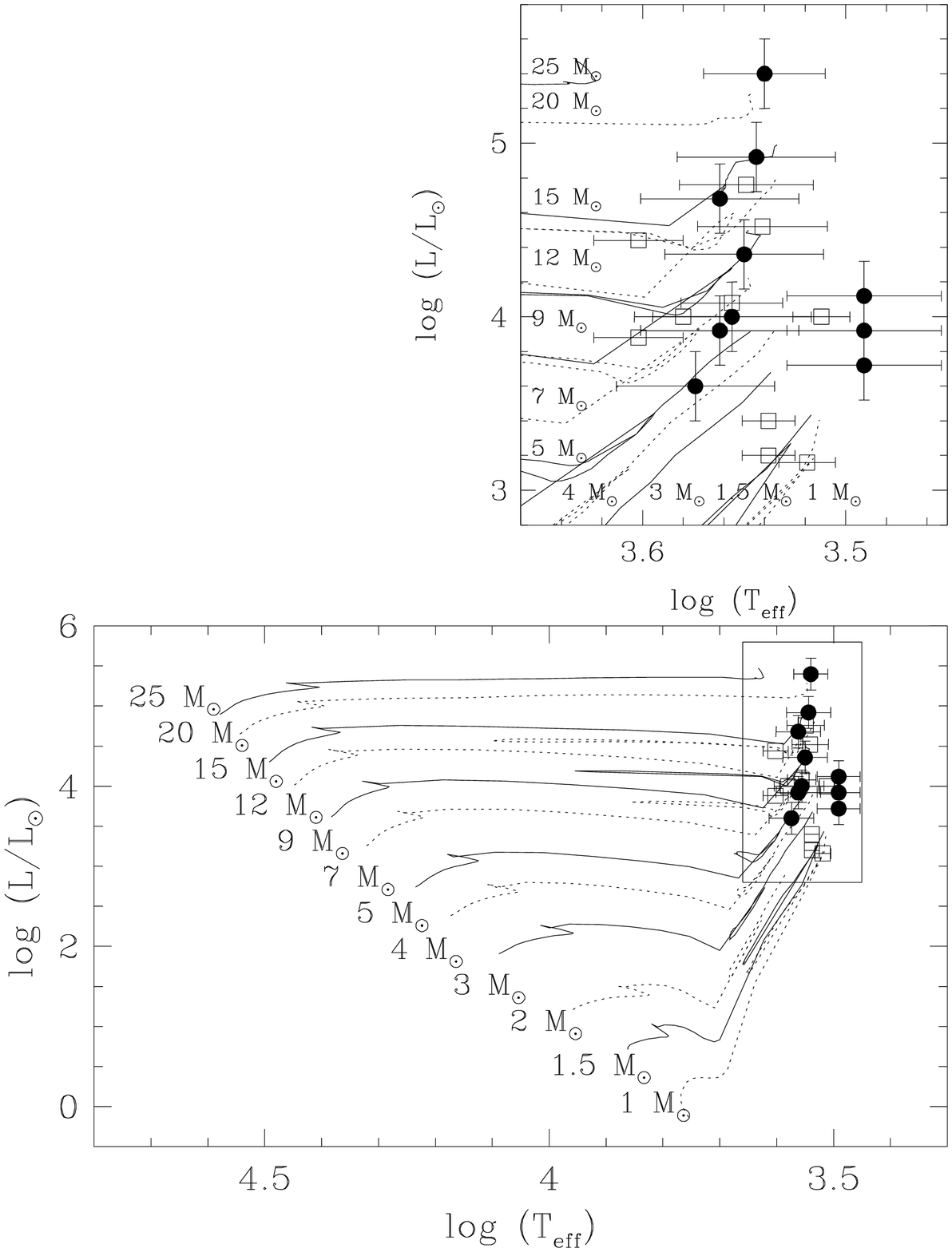}
\end{figure}

\begin{figure}
\epsfxsize \hsize
\epsfbox{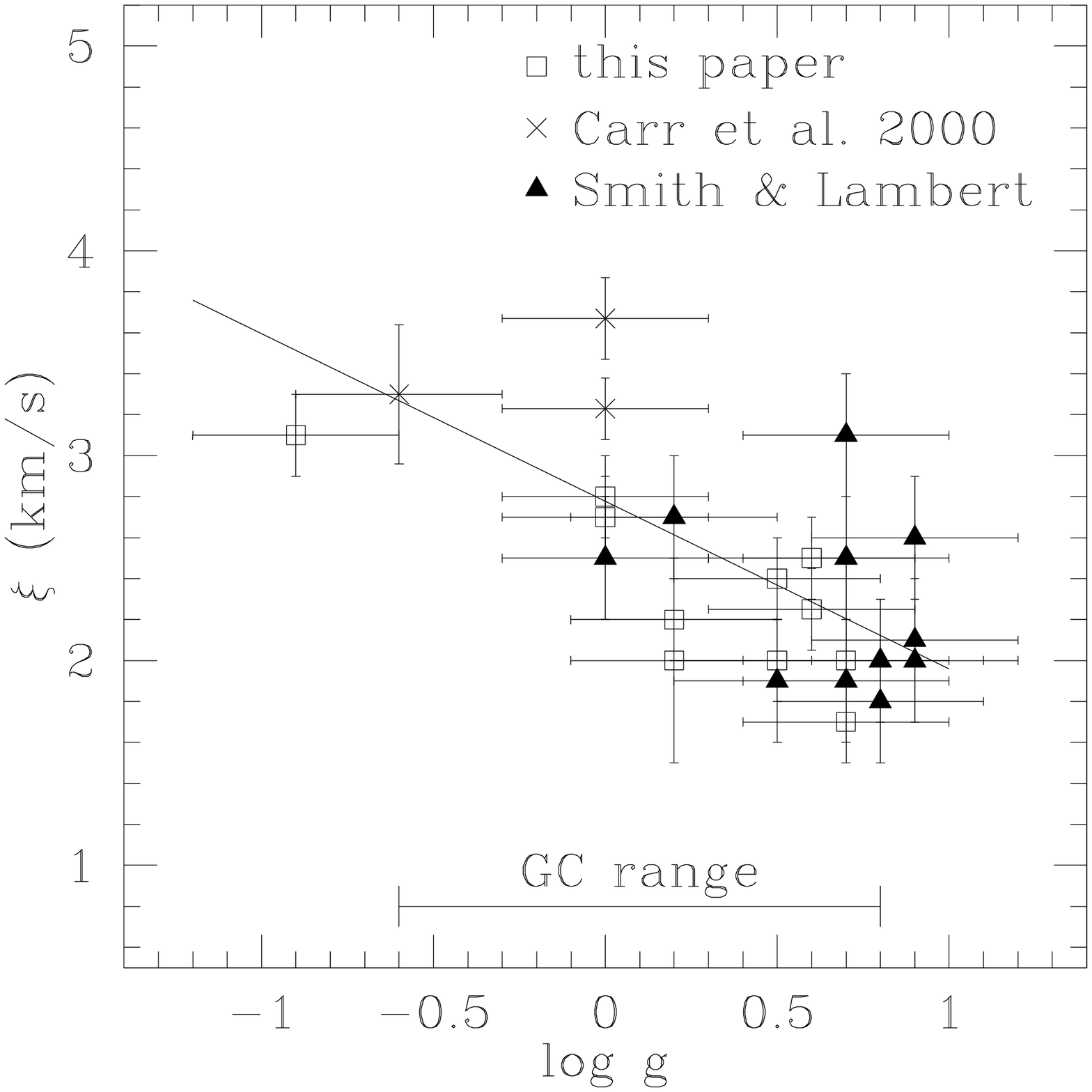}
\end{figure}

\begin{figure}
\epsfxsize \hsize
\epsfbox{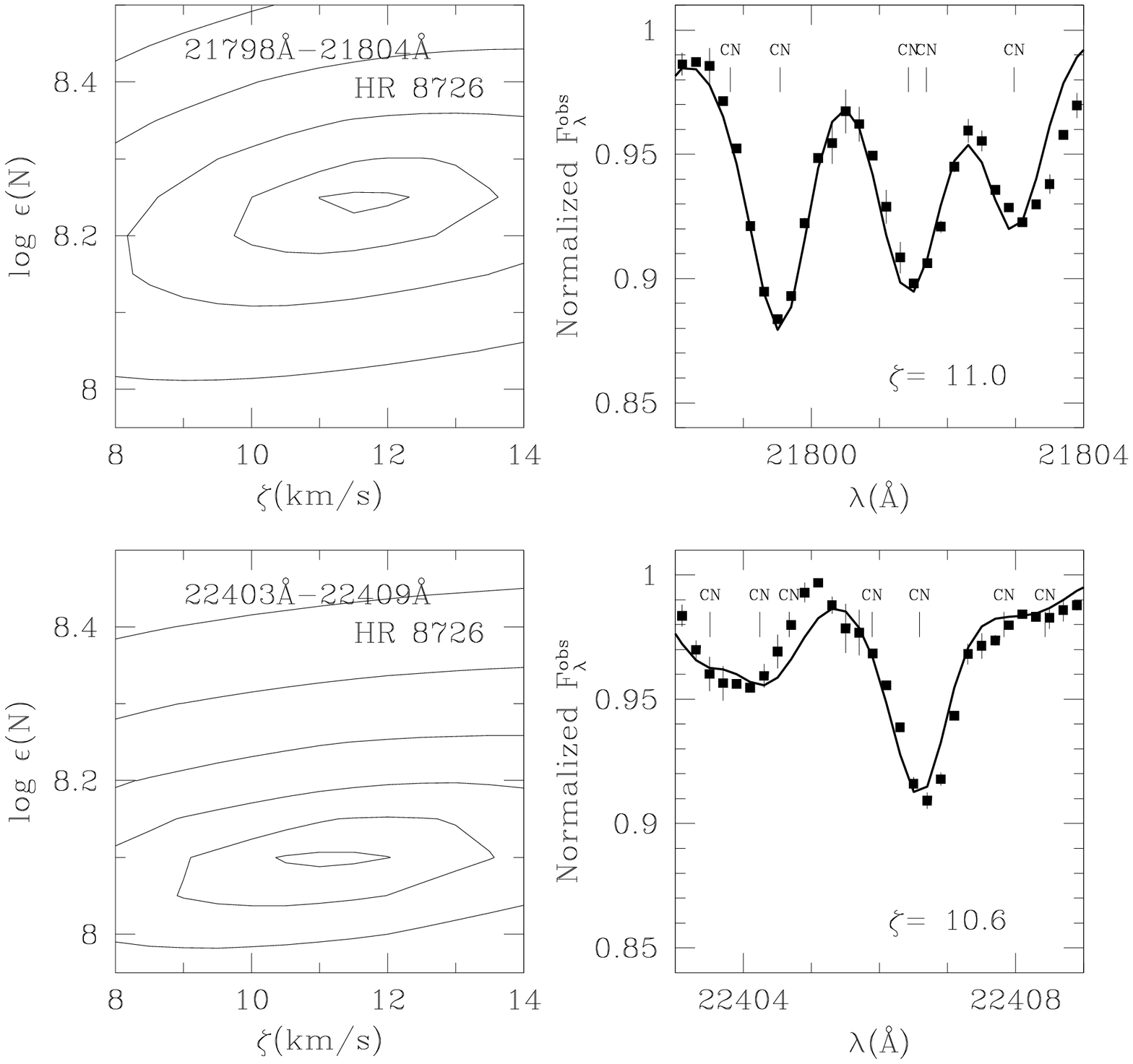}
\end{figure}

\begin{figure}
\epsfxsize \hsize
\epsfbox{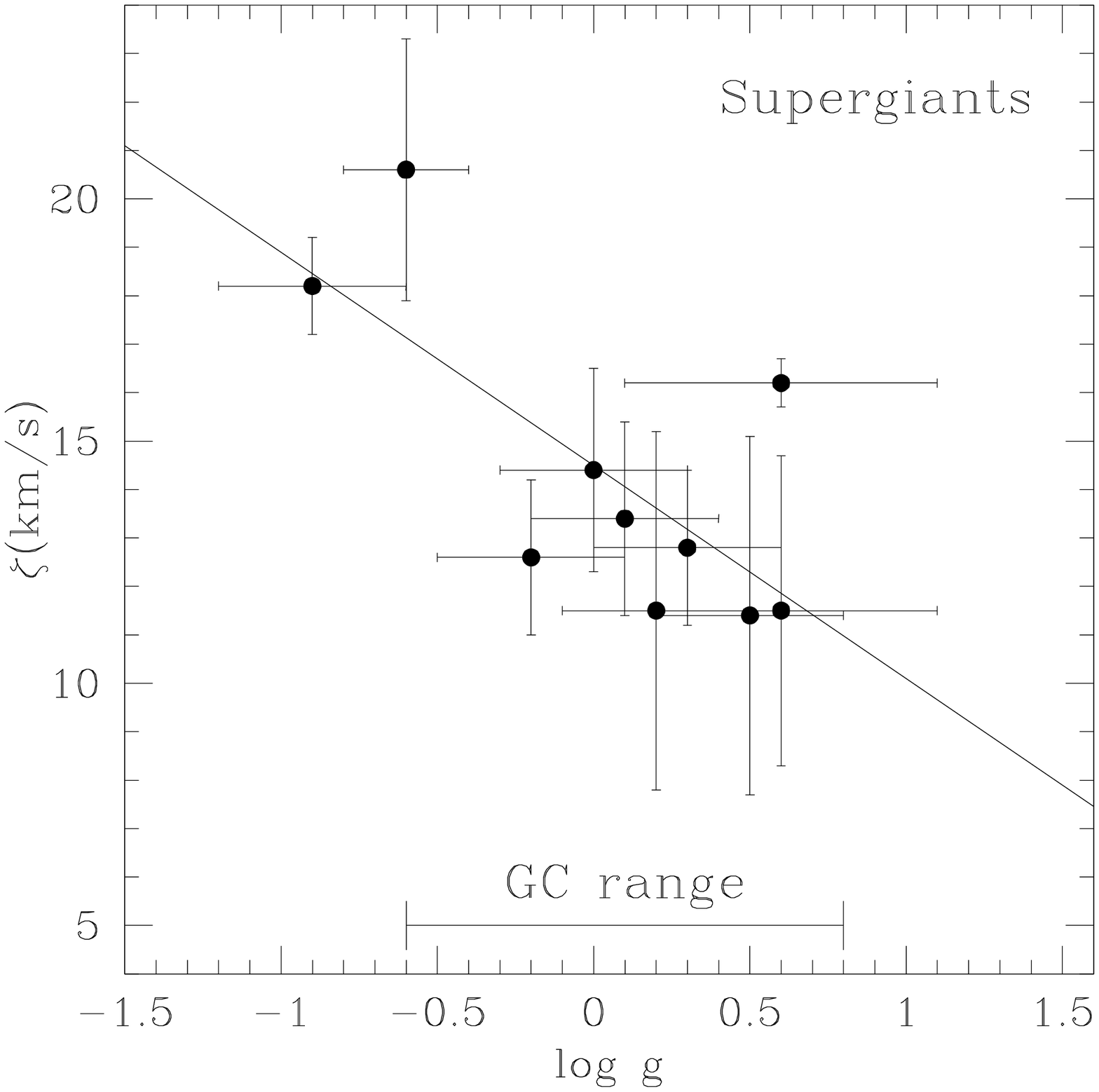}
\end{figure}

\begin{figure}
\epsfxsize \hsize
\epsfbox{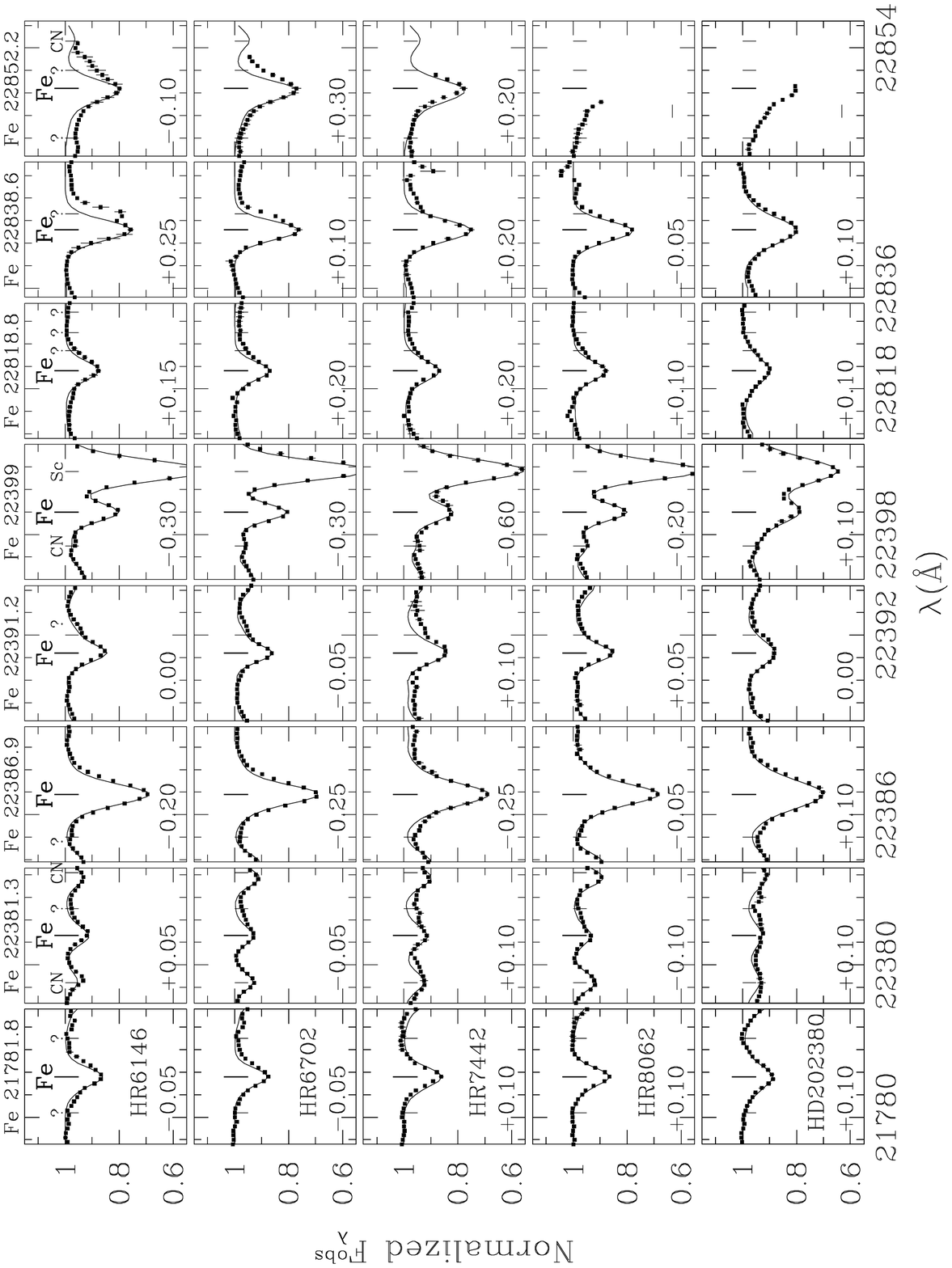}
\end{figure}

\begin{figure}
\epsfxsize \hsize
\epsfbox{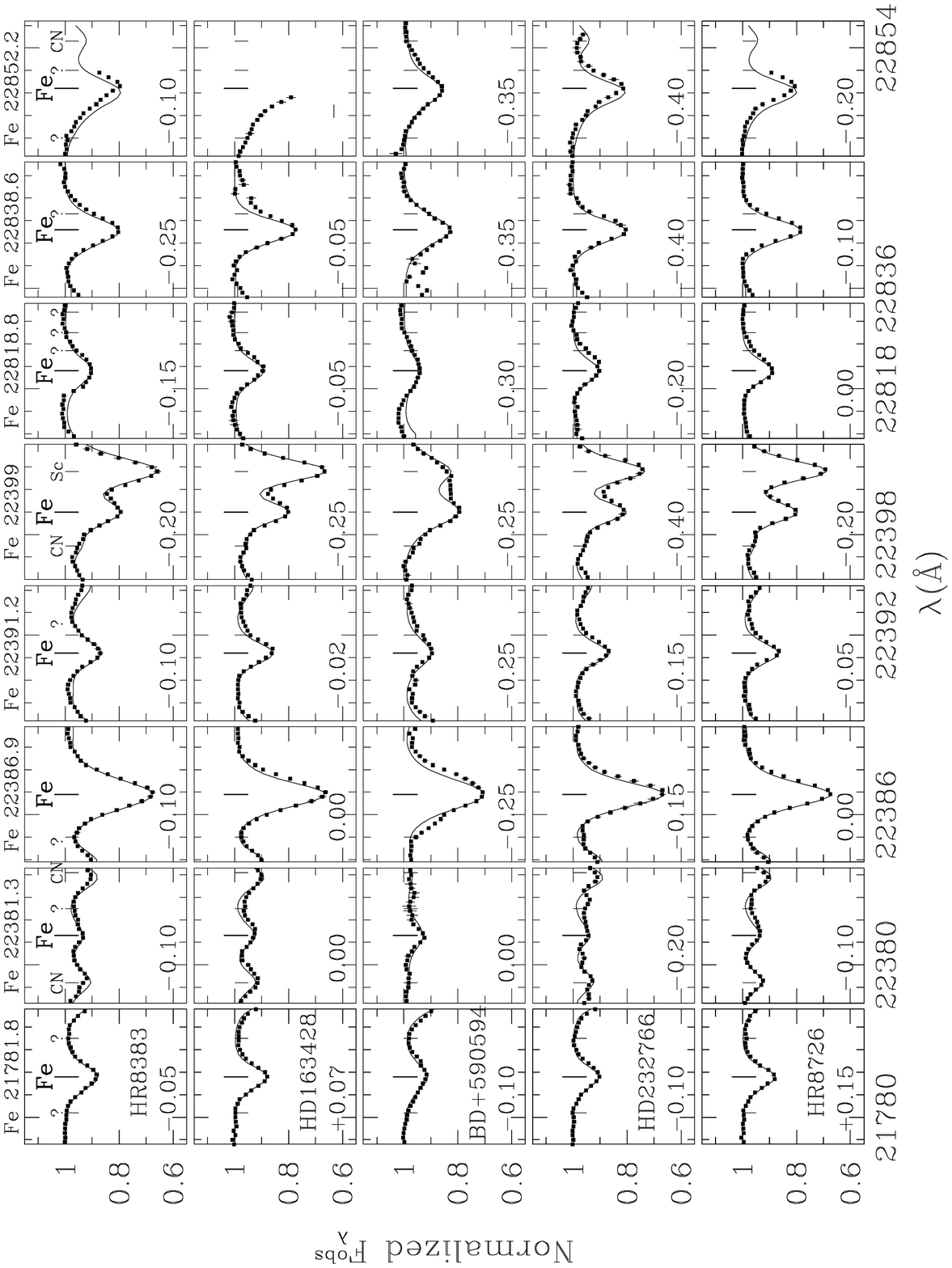}
\end{figure}

\begin{figure}
\epsfxsize \hsize
\epsfbox{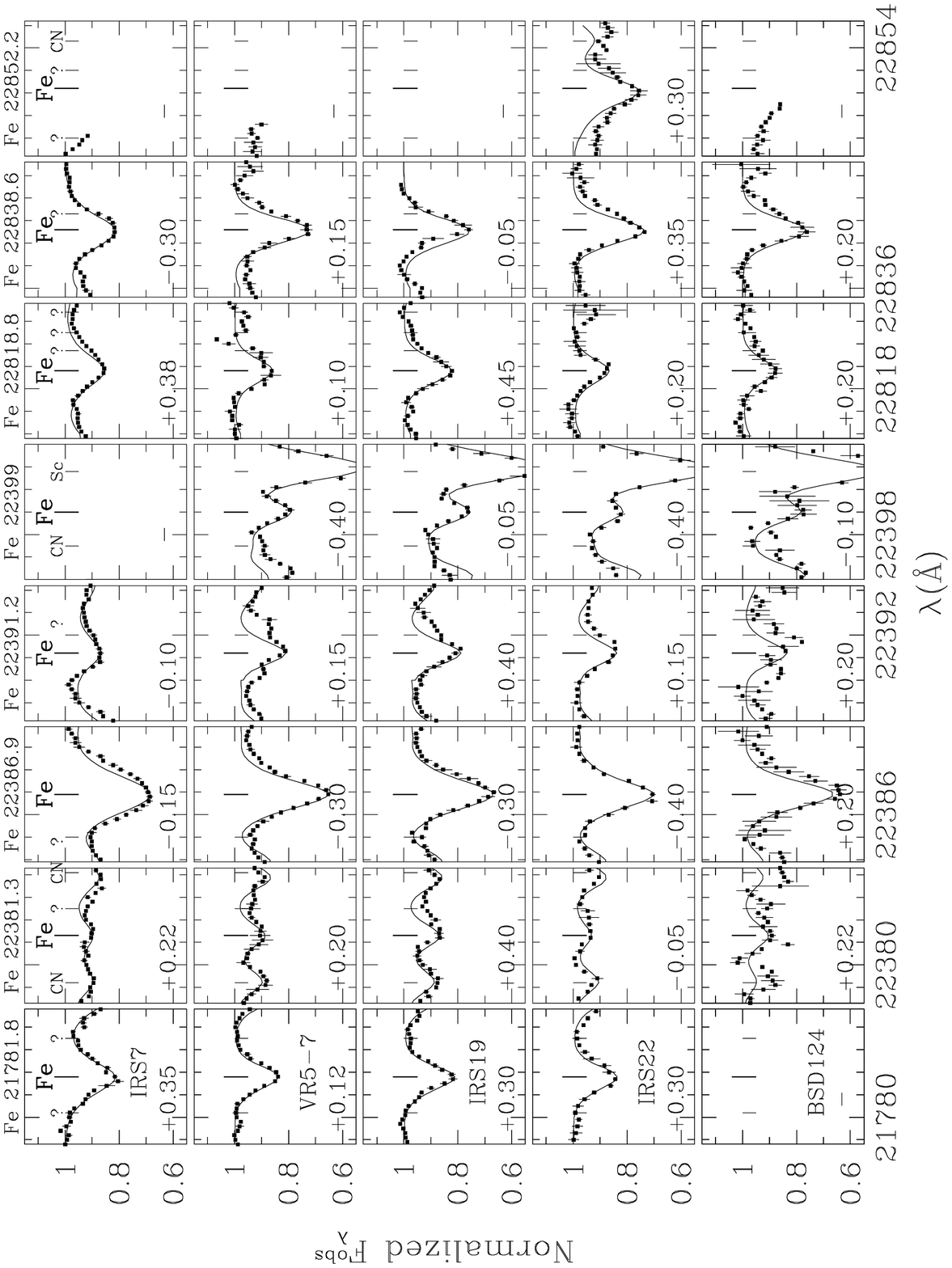}
\end{figure}

\begin{figure}
\epsfxsize \hsize
\epsfbox{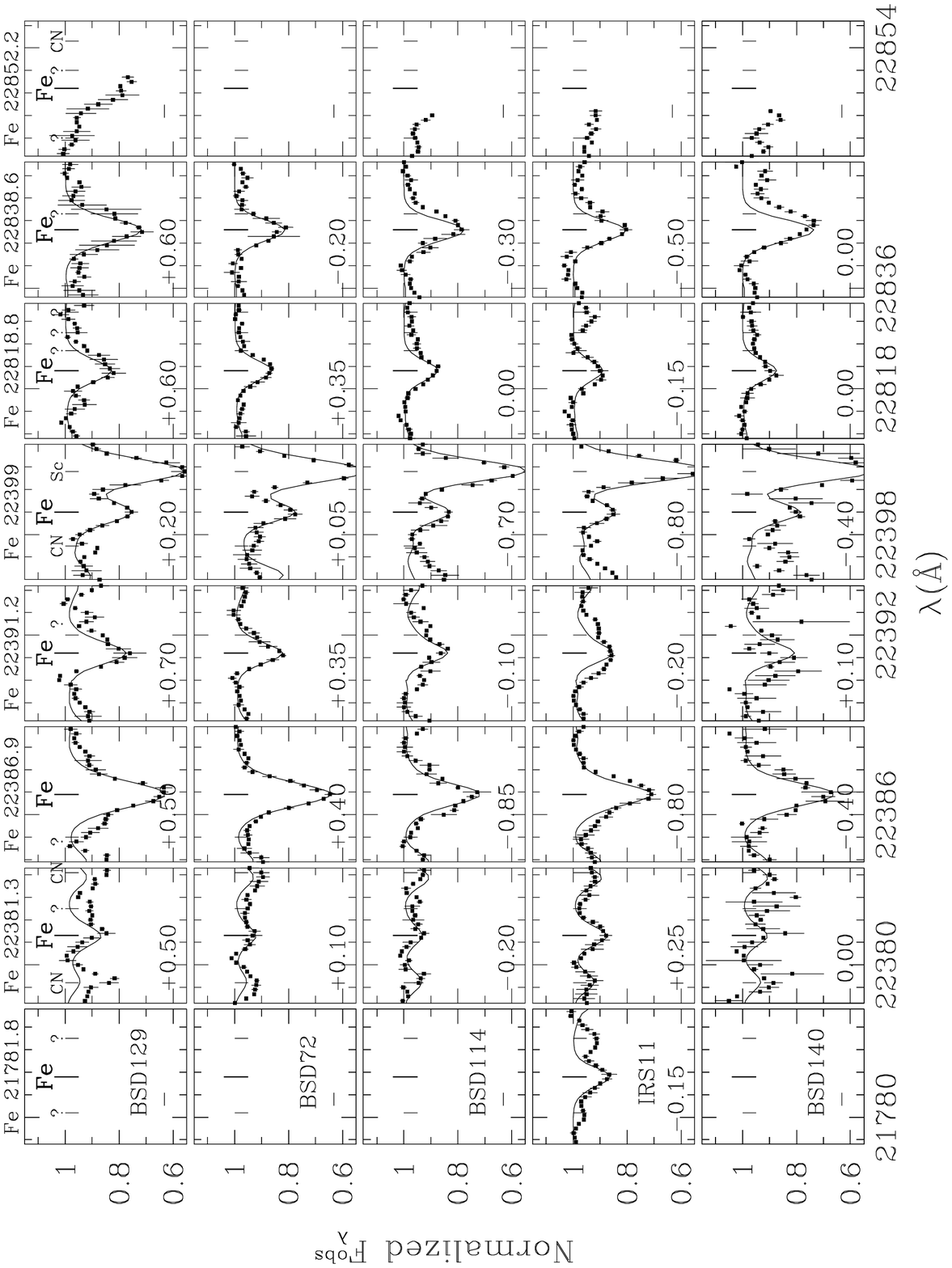}
\end{figure}

\begin{figure}
\epsfxsize \hsize
\epsfbox{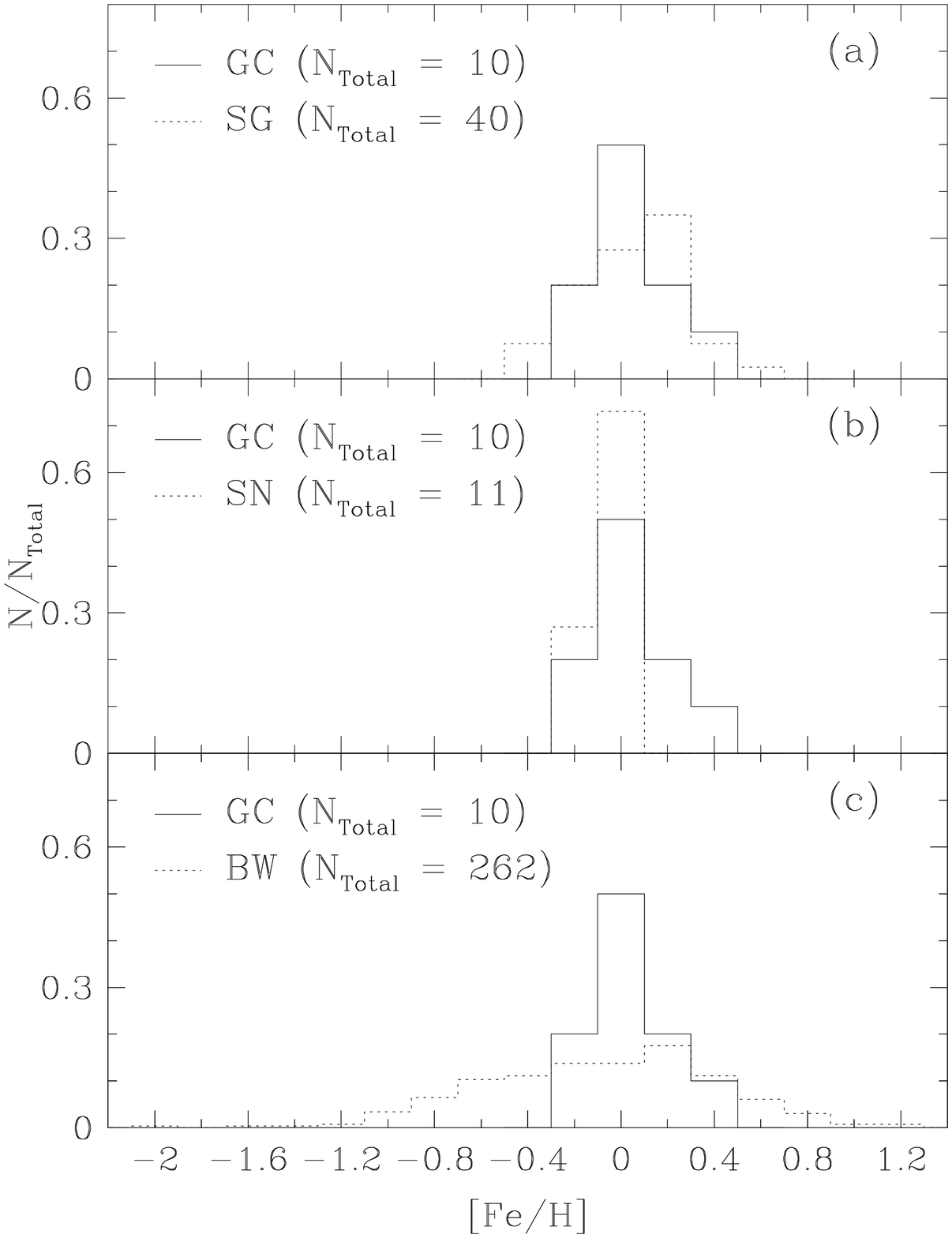}
\end{figure}

\end{document}